\documentclass[aps,prc,superscriptaddress,floats,floatfix,reprint,reprintnumbers]{revtex4}
\usepackage{amsmath,amssymb,bm,slashed}
\usepackage{booktabs}
\usepackage{blindtext}
\pdfoutput=1
\usepackage{epsfig}
\usepackage{graphicx}% Include figure files
\usepackage{dcolumn}% Align table columns on decimal point
\usepackage{bm}% bold math
\usepackage{booktabs}
\usepackage{color}
\usepackage{mathrsfs}
\usepackage{ulem}
\usepackage{epstopdf}
\graphicspath{{plots/}}% plot directory
\usepackage{verbatim}
\usepackage{indentfirst}
\usepackage{float}
\usepackage[bf,raggedright,indentafter]{titlesec}
\usepackage{rotating}
\usepackage{tabularx}
\usepackage{makecell}
\usepackage{subfigure}%
\usepackage{hyperref}

\usepackage{array}

\renewcommand{\baselinestretch}{1.3}

\begin{document}

\hspace{3cm}
\title{\large Majorana Neutrino Exchange in Free-Baryon $B_1B_2\to ppe^-e^-$ Reactions}
\author{Yuan-Guo Xu}
\email{yuanguoxu@jxnu.edu.cn}
\affiliation{School of Physics, Jiangxi Normal University, Nanchang, Jiangxi 330022, China}
\author{Bo Lan}
\email{lanbo@jxnu.edu.cn}
\affiliation{College of Physics and Electronic Information, Nanchang Normal University, Nanchang, Jiangxi 330032, China}
\author{Ben-Hou Xiang}
\email{xiangbh@ihep.ac.cn}
\affiliation{Institute of High Energy Physics, Beijing 100049, China}
\author{Shuang-Shi Fang}
\email{fangss@ihep.ac.cn}
\affiliation{Institute of High Energy Physics, Beijing 100049, China}
\author{Yi-Bing Zhang}
\email{yibingzhang@jxnu.edu.cn}
\affiliation{School of Physics, Jiangxi Normal University, Nanchang, Jiangxi 330022, China}
\author{Ru-Min Wang}
\email{ruminwang@sina.com}
\affiliation{School of Physics, Jiangxi Normal University, Nanchang, Jiangxi 330022, China}

\vspace{3cm}

\begin{abstract}
The lepton-number-violating free-baryon reactions $B_1B_2\to pp e^-e^-$ are studied for six nonstrange and strange initial states, with light- and heavy-Majorana-neutrino exchange treated in the same baryon-level framework. The calculation includes momentum-dependent weak form factors, all exchange terms associated with the identical final-state fermions, and exact four-body phase space. For light-neutrino exchange, the cross sections scale with the square of the effective Majorana mass $m_{\beta\beta}$. The calculated cross sections increase as the off-shell cutoff $\Delta q^2$ is reduced, especially near threshold and in the
hyperon channels, because more of the phase-space region where the squared four-momentum of the exchanged neutrino is close to zero is retained.
 At $\sqrt{s}=5.1~\mathrm{GeV}$, the six light-neutrino cross sections span $9.28\times10^{-49}$--$1.78\times10^{-45}~\mathrm{GeV}^{-2}$ for $m_{\beta\beta}=30~\mathrm{meV}$ and $\Delta q^2=10^{-2}~\mathrm{GeV}^2$. For heavy-neutrino exchange, the cross sections scale with $\eta_N^2$, where $\eta_N=|U_{eN}|^2/m_N$. The leading propagator correction remains below one percent for $m_N\gtrsim50~\mathrm{GeV}$ over the phase-space region considered. At $\sqrt{s}=5.1~\mathrm{GeV}$, the heavy-neutrino cross sections span $7.55\times10^{-45}$--$1.98\times10^{-41}~\mathrm{GeV}^{-2}$ for $\eta_N=4.0\times10^{-9}~\mathrm{GeV}^{-1}$. At higher energies, the two exchange cases give the same CKM-driven channel ordering, with the $nn$ cross section largest.
 At fixed $\sqrt{s}$, compared with the light-neutrino spectra calculated with $\Delta q^2=10^{-2}~\mbox{GeV}^2$, the heavy-neutrino peaks occur at a smaller squared invariant mass for the proton pair and a larger one
for the electron pair. These shifts remain after each spectrum is normalized to unit area. Thus, the two exchange cases differ in their spectral shapes, not only in their overall cross sections.
 These results extend the free-baryon analogue of neutrinoless double-beta decay to strange initial states and provide theoretical reference results for comparing light- and heavy-Majorana-neutrino exchange.

\end{abstract}
\maketitle

%\newpage
\section{Introduction}
\label{sec:introduction}

Neutrinoless double-beta decay ($0\nu\beta\beta$) is one of the most
sensitive laboratory probes of physics beyond the Standard Model. Its
observation would establish violation of total lepton number by two units,
$\Delta L=2$, and would imply that neutrinos have a Majorana mass component,
thereby addressing a central question raised by neutrino oscillations: the
origin and nature of neutrino mass
\cite{Vergados:2012xy,Rodejohann:2019,Agostini:2022zub}. In the standard
light-Majorana-neutrino exchange mechanism, the decay rate is controlled by
the effective Majorana mass $m_{\beta\beta}$, phase-space factors, and nuclear
matrix elements. However, the same experimental signature can also be generated
by a broad class of lepton-number-violating interactions, including heavy
sterile-neutrino exchange, right-handed currents, supersymmetric interactions,
and short-distance operators. Thus, $0\nu\beta\beta$ is sensitive to several
sources of lepton-number violation over a wide range of energy scales.

A major difficulty in relating present limits, or a future discovery, is the
uncertainty associated with hadronic and nuclear dynamics. Reliable nuclear
matrix elements are essential for translating half-life measurements into
constraints on particle-physics parameters. Recent progress in chiral effective
field theory has shown that even the light-neutrino mechanism requires a
leading-order short-range contribution
\cite{Cirigliano:2018,Cirigliano:2021LO}. Lattice-QCD and modern
nuclear-structure calculations are also beginning to provide additional input
\cite{Davoudi:2024,Yao:2022NME}. At the two-nucleon level, the
$nn\to pp e^-e^-$ transition has been studied as an input to nuclear
$0\nu\beta\beta$ calculations using lattice QCD and relativistic effective
field theory \cite{Davoudi:2024,YangZhao:2025PRL}. These studies address
low-energy nuclear amplitudes rather than free-baryon cross sections.

Experimentally, no confirmed signal has yet been observed, but the field has
entered a decisive phase. Searches using $^{76}$Ge, $^{130}$Te, $^{136}$Xe,
$^{100}$Mo, and other isotopes have pushed half-life limits to
$10^{25}$--$10^{26}$ years \cite{pdg2026}. Current and planned ton-scale
experiments aim to combine large isotope masses,
excellent energy resolution, and near-zero backgrounds, with sensitivity to
$m_{\beta\beta}$ in the inverted-ordering region and potentially beyond.
Interpreting a future signal will require observables that distinguish among
possible sources of lepton-number violation.

Free-baryon $\Delta L=2$ reactions remove nuclear bound-state wave functions
from the calculation and keep final-state momenta and angular correlations
explicit. They retain nontrivial dependence on baryonic transition form factors
and four-body kinematics.
Related \(\Delta L=2\) hyperon decays, \(B_i^-\to B_f^+\ell^-\ell'^-\), have been constrained experimentally \cite{LittenbergShrock:1992} and studied theoretically through Majorana-neutrino exchange, short-range four-quark operators, and covariant baryon chiral perturbation theory \cite{Barbero:2003Hyperon,Barbero:2013Hyperon,HernandezTome:2022Hyperon,Zhao:2026klv}. Unlike these one-baryon decays, the present work considers free two-baryon scattering constructed from products of single-baryon charged currents.

In this work, the initial and final baryons are described by free on-shell spinors, and the amplitudes are built from products of single-baryon charged currents connected by the exchanged neutrino.
Strong-interaction effects within each baryon transition are included through the form factors. Strong interactions between the two initial baryons and between the two final protons, Coulomb interactions among
the charged final particles, and direct two-baryon short-distance contributions are not included \cite{Cirigliano:2018,Cirigliano:2021LO}. Including both $d\to u$ and $s\to u$ charged-current transitions gives six channels, whose CKM factors, thresholds, transition form factors, and four-body phase space can be compared consistently.  The calculation examines their ordering at higher energies, cutoff dependence, and the distributions in the squared invariant masses of the proton and electron pairs.

The paper is organized as follows.  Section~\ref{sec:framework} introduces the effective weak interaction, constructs the antisymmetrized amplitudes, and derives the four-body phase-space representation and the prescriptions for
light- and heavy-neutrino exchange.  The numerical results for the total and differential cross sections are given in Sec.~\ref{sec:numerical}.  The final section summarizes the main results.

\section{Theoretical framework}
\label{sec:framework}
The lepton-number-violating scattering considered here is
\begin{equation}
B_1(p_1)+B_2(p_2)\to
p(p_3)+p(p_4)+e^-(k_1)+e^-(k_2),
\label{eq:process}
\end{equation}
where $B_1B_2=nn,\ n\Lambda,\ \Lambda\Lambda,\ n\Sigma^0,\ \Sigma^0\Sigma^0,\ \Lambda\Sigma^0$. Here, $B_1$ and $B_2$ are the initial baryons, while $p$ and $e^-$ denote the final protons and electrons.
The symbols $p_{1,2,3,4}$ and $k_{1,2}$ denote the baryon and electron four-momenta, respectively.  The light- and heavy-Majorana-neutrino exchange amplitudes are constructed following the standard \(0\nu\beta\beta\) formalism \cite{Doi:1985dx}.

\subsection{Weak interaction and  cross section}
At energies well below the electroweak scale, the charged-current interaction at a baryonic vertex is written as
\begin{equation}
{\cal H}_{W}^{(q)}=\frac{G_F}{\sqrt{2}}\,V_{uq}\,H_{Bp}^{\mu}\,\bar e\,\gamma_\mu(1-\gamma_5)\nu_e+\mathrm{H.c.} ~~~\mbox{with}~~~q=d,s,
\label{eq:weakHamiltonian}
\end{equation}
where $G_F$ is the Fermi constant, $H_{Bp}^{\mu}$ is the baryonic transition current, $\nu_e$ is the electron-neutrino field, and H.c. denotes the Hermitian conjugate. The CKM matrix element $V_{uq}$ is $V_{ud}$ for a $d\to u$ transition and $V_{us}$ for an $s\to u$ transition.
The hadronic current is parameterized by
\begin{equation}
H_{Bp}^{\mu}(p',p)=\bar u_p(p')\gamma^\mu\left[g_V^{Bp}(t)-g_A^{Bp}(t)\gamma_5\right]u_B(p),
\label{eq:hadronicCurrent}
\end{equation}
where $u_B(p)$ and $u_p(p')$ are the Dirac spinors of the initial baryon and final proton, respectively, $t=(p-p')^2$ is the squared momentum transfer, and the form factors, $g_V^{Bp}$ and $g_A^{Bp}$,
are given in Appendix~\ref{AppendixA}, using commonly adopted parametrizations of neutron and hyperon semileptonic transitions \cite{Simkovic:1999,Cabibbo:2003,Ratcliffe:2004jt,Mateu:2005wi}.  Only the leading vector and axial-vector currents are retained. Weak-magnetism, induced-scalar, induced-tensor, and induced-pseudoscalar terms are neglected. The resulting absolute cross sections therefore depend on this hadronic-current
approximation.

The final state contains two identical protons and two identical electrons, so the amplitude must be antisymmetrized with respect to both pairs
\begin{equation}
{\cal M}= {\cal M}_{34;12}-{\cal M}_{43;12} -{\cal M}_{34;21} +{\cal M}_{43;21},\label{eq:antisymM}
\end{equation}
with
\begin{align}
{\cal M}_{ab;ij} ={}& \frac{G_F^2}{2}\,V_{uq} V_{uq'}
H_{B_1p}^{\mu}(p_a,p_1)\,H_{B_2p}^{\nu}(p_b,p_2) \notag\\
&\times L_{\mu\nu}(k_i,k_j;q_{ai}),
\label{eq:Mabij}
\end{align}
where $(a,b)=(3,4)$ or $(4,3)$, $(i,j)=(1,2)$ or $(2,1)$, while $q,q'=d,s$ label the quark flavors at the two weak vertices. The relative signs in Eq.~\eqref{eq:antisymM} follow from fermion exchange.
In $\mathcal M_{ab;ij}$, the internal Majorana-neutrino propagator connecting the two charged-current vertices carries four-momentum $q_{ai}\equiv p_1-p_a-k_i$, with $a=3,4$ and $i=1,2$.

The leptonic part of the amplitude can be written as
{\small
\begin{align}
L_{\mu\nu}(k_i,k_j;q_{ai})
={}&
\sum_{i'} U_{ei'}^2\,
\bar u(k_i)\gamma_\mu(1-\gamma_5) \notag\\
&\times
\frac{i(\slashed q_{ai}+m_{\nu_{i'}})}
     {q_{ai}^2-m_{\nu_{i'}}^2+i0^+}
(1-\gamma_5)\gamma_\nu
C\bar u^{\,T}(k_j),
\label{eq:leptonicChain}
\end{align}}
where $U_{ei'}$ are elements of the first row of the Pontecorvo-Maki-Nakagawa-Sakata matrix \cite{Pontecorvo:1957qd,Maki:1962mu}. The electron-neutrino field is
$\nu_e=\sum_{i'}U_{ei'}\nu_{i'}$, where $m_{\nu_{i'}}$ is the mass of the neutrino eigenstate $\nu_{i'}$. Here, $C$ is the charge-conjugation matrix, and
$C\bar u^{\,T}(k_j)$ is the charge-conjugate electron spinor~\cite{Zhao:2026klv}.

After summing over final spins and averaging over initial spins, the squared amplitude is
{\small
\begin{align}
\overline{|{\cal M}|^2}=\frac{G_F^4 |V_{uq}|^2 |V_{uq'}|^2}{16}\Big(T_{11}-T_{12}-T_{13}+T_{14} -T_{21}+T_{22}+T_{23}-T_{24} -T_{31}+T_{32}+T_{33}-T_{34} +T_{41}-T_{42}-T_{43}+T_{44}\Big),
\label{eq:M2all}
\end{align}}\noindent
where $T_{mn}$ is the spin-summed product of the $m$th and complex-conjugate $n$th terms in Eq.~\eqref{eq:antisymM}, after removing their common factor $G_F^2V_{uq}V_{uq'}/2$.
Eq.~\eqref{eq:M2all} already contains the factor $1/4$ from the initial-spin average.  An example of $T_{mn}$ is given in  Appendix \ref{AppendixA}.

Using the kinematic analysis and the four-body phase space given in Appendix \ref{AppendixA}, the differential cross section  is
\begin{align}
d\sigma ={}& \frac{Y\,Y_{pp}\,Y_{ee}\,\overline{|{\cal M}|^2}}
{2^{16}\pi^7\sqrt{s}\,\sqrt{\lambda(s,m_{B_1}^2,m_{B_2}^2)}\,m_{pp}m_{ee}}
\,dm_{pp}^2\,dm_{ee}^2 \notag\\
&\times d\cos\theta_{\rm CM}\,d\cos\theta_{pp}\,d\phi_{pp}
\,d\cos\theta_{ee}\,d\phi_{ee},
\label{eq:dsigma7D}
\end{align}
where $Y$, $Y_{pp}$, $Y_{ee}$, and the five angles are defined in Appendix~\ref{AppendixA}.
The seven-dimensional integral is evaluated by quasi-Monte-Carlo
integration with up to $10^7$ sampling points for each channel,
energy, and cutoff. The integration is performed directly over the
physical limits given in Appendix~\ref{AppendixA}, with the off-shell
condition imposed in the integrand. For the $nn$ channel at
$\sqrt{s}=2.1~\mathrm{GeV}$ and the reference cutoff
$\Delta q^2=10^{-2}~\mathrm{GeV}^2$, increasing the maximum number of
sampling points from $10^5$ to $10^6$ and then to $10^7$ changes the
result by $0.32\%$ and $0.083\%$, respectively. At
$\Delta q^2=10^{-4}~\mathrm{GeV}^2$, convergence is slower because of
the narrow region near $q_{ai}^2=0$, producing the visible numerical
variations in several curves.

\subsection{Off-shell prescription}
Light and heavy Majorana neutrinos require different treatments of the internal propagator.

\subsubsection{Light-Majorana-neutrino contribution}
In the light-neutrino limit, $m_{\nu_{i'}}^2\ll |q_{ai}^2|$, the neutrino factor in the amplitude reduces to
\begin{equation}
\sum_{i'} U_{ei'}^2\frac{m_{\nu_{i'}}}{q_{ai}^2-m_{\nu_{i'}}^2+i0^+}\simeq\frac{1}{q_{ai}^2}\sum_{i'}U_{ei'}^2m_{\nu_{i'}}.
\label{eq:propagator_approx}
\end{equation}
The effective Majorana mass is defined as $m_{\beta\beta}=\left|\sum_{i'}U_{ei'}^2m_{\nu_{i'}}\right|$, as in conventional $0\nu\beta\beta$ decay
\cite{Doi:1985dx,Avignone:2007fu,Rodejohann:2019,Agostini:2022zub}.
For light-neutrino exchange alone, the squared amplitude is therefore proportional to $m_{\beta\beta}^{2}$. This factor includes the interference between different
neutrino mass eigenstates and is used directly in the numerical calculation.

Within the four-body phase space, $q_{ai}^2$ can approach zero from the spacelike region, where the light-neutrino propagator $1/q_{ai}^2$ becomes singular. We therefore require
\begin{equation}
q_{ai}^2-m_{\nu_{i'}}^2<-\Delta q^2~~~~ \mbox{with} ~~~~q_{ai}=q_{31},q_{41},q_{32},q_{42}.
\label{eq:cutgeneral}
\end{equation}
The cutoff is applied separately to each exchange amplitude before
the four terms are summed coherently. It is therefore an
amplitude-level phenomenological prescription rather than an event
selection and may modify the interference terms. The calculation
excludes all timelike configurations and gives only the spacelike
contribution defined by this prescription, not the complete physical
cross section.

In the numerical light-neutrino limit, $m_{\nu_{i'}}\to0$ is set in
the denominator, while $m_{\beta\beta}$ is retained in the numerator.
Equation~\eqref{eq:cutgeneral} then becomes
$q_{ai}^2<-\Delta q^2$. Accordingly, the quoted light-neutrino cross sections are the spacelike contributions defined by this cutoff. The following values are used:
\begin{equation}
\Delta q^2=10^{-4},\ 10^{-3},\ 3\times10^{-3},\ 10^{-2},\ 3\times10^{-2},\ 10^{-1} \ {\rm GeV}^2 .
\label{eq:Deltalist}
\end{equation}

For fixed $\sqrt{s}$, cutoff, and other input parameters, the light-neutrino-exchange amplitude is proportional to the effective Majorana mass $m_{\beta\beta}$; therefore, the cross section scales as $\sigma\propto m_{\beta\beta}^{\,2}$.  The following values are taken in this work
\begin{equation}
m_{\beta\beta}=1,\ 10,\ 30,\ 50,\ 100~{\rm meV}.
\end{equation}
 They are representative points and do not define an experimentally allowed interval. The complete KamLAND-Zen data set gives $m_{\beta\beta}<28$--$122$~meV at 90\% C.L., and this range of upper limits reflects the
choice of nuclear matrix elements \cite{KamLANDZen:2025}.

\subsubsection{Heavy-Majorana-neutrino contribution}

For a heavy Majorana mass eigenstate $N$, the propagator between the two left-handed charged-current vertices is proportional to
\begin{equation}
 U_{eN}^{\,2} \frac{m_N} {q_{ai}^2-m_N^2+i0^+},
 \label{eq:heavy-propagator}
\end{equation}
where $U_{eN}$ denotes the mixing between the electron neutrino and the heavy Majorana state $N$, and $m_N$ is the corresponding Majorana mass.
In the heavy-neutrino regime ($m_N^2 \gg |q_{ai}^2|$), the propagator can be expanded as
\begin{equation}
 U_{eN}^{\,2}  \frac{m_N}  {q_{ai}^2-m_N^2+i0^+}  = -\frac{U_{eN}^{\,2}}{m_N} \left[  1+\mathcal{O}\left(\frac{q_{ai}^2}{m_N^2}\right) \right].
 \label{eq:heavy-propagator-expansion}
\end{equation}
The spin-averaged squared amplitude therefore scales as
\begin{equation}
 \overline{|\mathcal M_N|^2} \propto G_F^4 |V_{uq}|^2 |V_{uq'}|^2 \eta_N^2  ~~~~\mbox{with}~~~~ \eta_N  \equiv  \frac{|U_{eN}|^2}{m_N},
 \label{eq:M2-heavy}
\end{equation}
where $\mathcal M_N$ denotes the heavy-neutrino-exchange amplitude.
Within this approximation, the cross section depends on $m_N$ and $|U_{eN}|$ only through $\eta_N$ \cite{Faessler:2014kka,Dekens:2020Sterile,Atre:2009Heavy}. This work considers one heavy Majorana state. If more than one heavy
state contributes, the amplitude depends on $\sum_N U_{eN}^2/m_N$, and the cross section depends on the absolute square of this sum.

Limits on this combination can be inferred from  $0\nu\beta\beta$ decay searches if heavy-neutrino exchange is assumed to be the dominant mechanism.
Using the coefficients $\mathcal A$ in Tab.~I and the experimental half-life lower bounds in Eq.~(18) of Ref.~\cite{Faessler:2014kka},
the following upper limits are obtained in the heavy-neutrino limit
\begin{eqnarray}
 \eta_N  \big({}^{136}{\rm Xe}\big)&<& (4.12\text{--}8.70)\times10^{-9}\ {\rm GeV}^{-1}, \label{eq:etaN-Xe}\\
 \eta_N\big({}^{76}{\rm Ge}\big) &<& (5.74\text{--}12.2)\times10^{-9}\ {\rm GeV}^{-1}, \label{eq:etaN-Ge}
\end{eqnarray}
where this range mainly reflects the nuclear-model dependence of the matrix elements. Two representative parameter values are therefore adopted:
\begin{equation}
 \eta_N^{A} = 4.0\times10^{-9}\ {\rm GeV}^{-1}, \qquad \eta_N^{B} = 8.0\times10^{-9}\ {\rm GeV}^{-1}.
 \label{eq:etaN-values}
\end{equation}
These two values are used as representative normalizations based on the range inferred from the ${}^{136}\mathrm{Xe}$ results and nuclear matrix elements considered in Ref.~\cite{Faessler:2014kka}. They
should not be interpreted as updated limits derived from the recent KamLAND-Zen result~\cite{KamLANDZen:2025}. Direct-search limits constrain $m_N$ and $|U_{eN}|^2$ separately and depend on the assumed
heavy-neutrino mass and mixing pattern; they therefore cannot be expressed as a single mass-independent limit on $\eta_N$. Since the cross sections scale as $\eta_N^2$ in the heavy-neutrino limit, the
results can be rescaled to any chosen value of $\eta_N$.

The heavy-neutrino approximation requires $m_N^2\gg|q_{ai}^2|$ over the full physical phase space. The four squared neutrino momenta $q_{31}^2$, $q_{32}^2$, $q_{41}^2$, and $q_{42}^2$ are therefore examined
for all six channels and for the CM energies up to $\sqrt{s}=5.1$~GeV.  Their maximum magnitude is defined by
\begin{equation}
 Q_{\rm max}^2 = \max_{\rm channels,\;phase\;space} \left\{ |q_{31}^2|,~ |q_{32}^2|,~ |q_{41}^2|,~ |q_{42}^2| \right\},
 \label{eq:Qmax-heavy}
\end{equation}
and a numerical calculation gives $Q_{\rm max}^2 \simeq 22.45~{\rm GeV}^2$. Requiring $Q_{\max}^2/m_N^2<0.01$ keeps the leading relative correction in Eq.~(\ref{eq:heavy-propagator-expansion}) below one percent and gives
$m_N>\sqrt{Q_{\max}^2/0.01}\simeq 47.5~\mathrm{GeV}$. This criterion concerns the propagator expansion, not the full theoretical accuracy of the cross section.
Thus, for $m_N\gtrsim50~\mathrm{GeV}$, the leading propagator correction remains below one percent over the phase-space region considered.

For example, the value $\eta_N^A$ may be represented by
\begin{align}
 (m_N,|U_{eN}|)={}&(100~{\rm GeV},\,6.32\times10^{-4}), \notag\\
 &(1000~{\rm GeV},\,2.00\times10^{-3}),\ldots
 \label{eq:etaNA-examples}
\end{align}
whereas $\eta_N^B$ corresponds, for example, to
\begin{align}
 (m_N,|U_{eN}|)={}&(100~{\rm GeV},\,8.94\times10^{-4}), \notag\\
 &(1000~{\rm GeV},\,2.83\times10^{-3}),\ldots
 \label{eq:etaNB-examples}
\end{align}
Within the heavy-neutrino limit, each pair gives the same result for its fixed $\eta_N$.

In this heavy-neutrino-mass regime, no spacelike cutoff $\Delta q^2$ is needed. For masses satisfying $m_N>50~\mathrm{GeV}$,
the pole at $q_{ai}^2=m_N^2$ lies far outside the accessible range of $q_{ai}^2$. The exchanged neutrino therefore remains far off shell throughout the integration.

\section{Numerical results and discussion}
\label{sec:numerical}

The particle masses and the Fermi constant are taken from the Particle Data Group \cite{pdg2026}. The CKM values $|V_{ud}|=0.97433$ and $|V_{us}|=0.22507$ are taken from the matrix displayed by the UTfit Collaboration \cite{UTfit}. Central input values are used, and their uncertainties are not included.

\subsection{Light-neutrino exchange}
For light-Majorana-neutrino exchange, the cutoff and energy dependences are evaluated at $m_{\beta\beta}=30~\mathrm{meV}$. Results for other values of $m_{\beta\beta}$ follow from the quadratic scaling given below.

Fig.~\ref{fig:LN-total} shows the  cross sections for light-Majorana-neutrino exchange as functions of the CM energy $\sqrt{s}$ for all six channels  with
$m_{\beta\beta}=30~\mathrm{meV}$.
\begin{figure}[t]
\begin{center}
\includegraphics[scale=0.88]{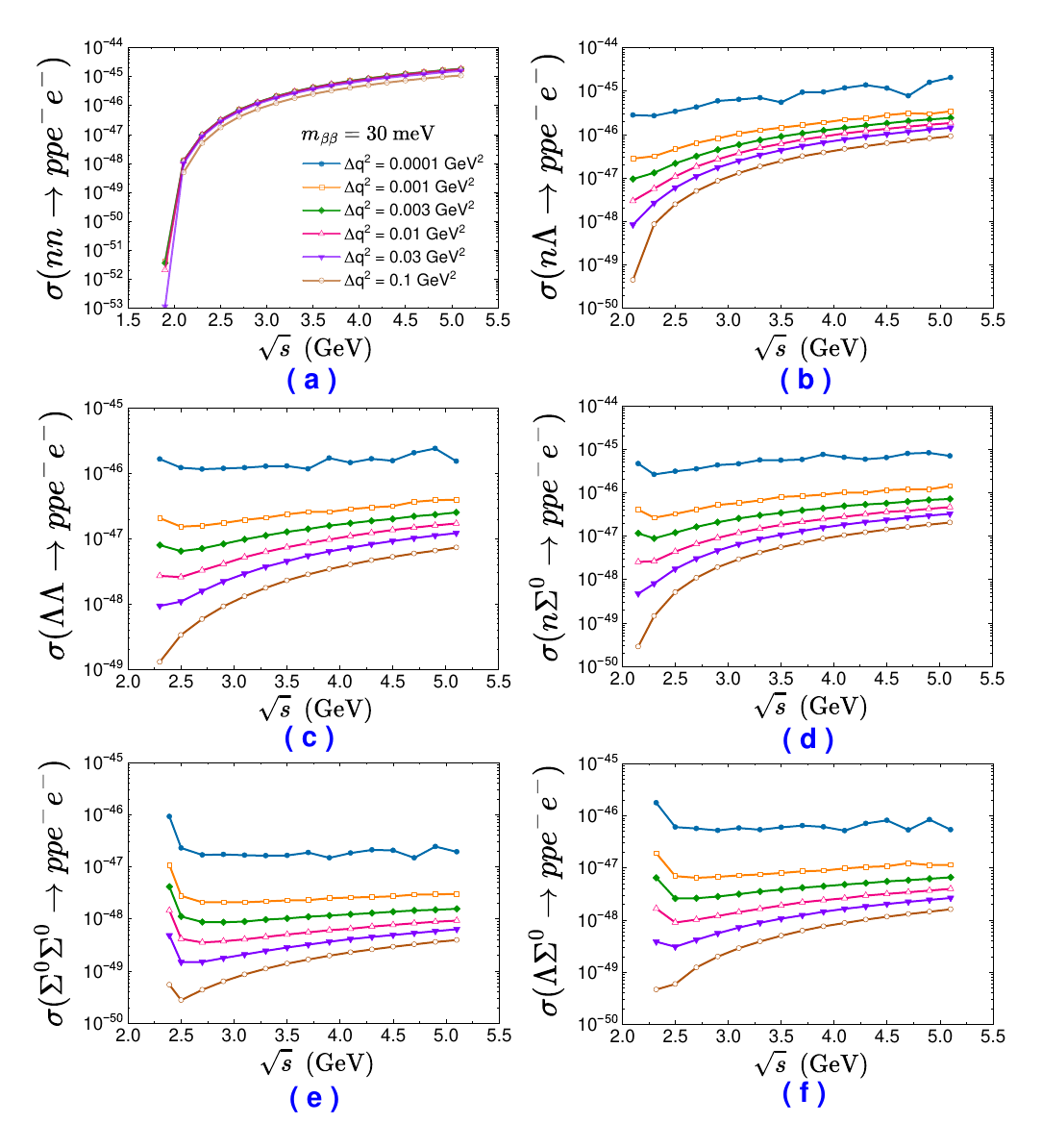}
\end{center}
\caption{Total cross sections for the light-Majorana-neutrino contribution as functions of the CM energy $\sqrt{s}$, with $m_{\beta\beta}=30~\mathrm{meV}$. The six curves correspond to $\Delta q^2=10^{-4},\,10^{-3},\,3\times10^{-3},\,10^{-2},\,3\times10^{-2}$, and $10^{-1}~\mathrm{GeV}^2$. Panels (a)-(f) show $nn$, $n\Lambda$, $\Lambda\Lambda$, $n\Sigma^0$, $\Sigma^0\Sigma^0$, and $\Lambda\Sigma^0\to pp e^-e^-$, respectively. The cross sections are given in $\mathrm{GeV}^{-2}$.}
\label{fig:LN-total}
\end{figure}
The $nn$ cross section rises rapidly near threshold, whereas the hyperon channels show a stronger cutoff dependence. For smaller cutoffs, the hyperon cross sections remain nearly flat or decrease at first before rising. This feature becomes weaker or disappears as the cutoff increases. At higher energies, the cross sections generally increase with $\sqrt{s}$, with only small numerical variations at the smallest cutoff.

 Very close to the incoming threshold, the initial-state flux factor in Eq.~\eqref{eq:dsigma7D} tends to zero. Since the initial-baryon masses exceed the total final-state mass in all six channels, the cross section may vary inversely with the initial relative velocity. The threshold trends therefore depend on the flux factor as well as phase space, form factors, and the neutrino propagator.

Besides changing the near-threshold behavior, the cutoff also
affects the size of the cross sections. Across the six cutoff
values in Fig.~\ref{fig:LN-total}, a smaller $\Delta q^2$ retains
more phase space near $q_{ai}^2=0$, where the light-neutrino
propagator enhances the amplitude. The plotted cross sections
increase as $\Delta q^2$ decreases. At fixed $\sqrt{s}$, apart
from small numerical variations, they satisfy
{\small
\begin{align}
\sigma(10^{-4})&>\sigma(10^{-3})>\sigma(3\times10^{-3}) \notag>\sigma(10^{-2})>\sigma(3\times10^{-2})>\sigma(10^{-1}),
\end{align}}
where the values of $\Delta q^2$ are in units of $\mathrm{GeV}^{2}$.
For $\Delta q^2=10^{-4}~\mathrm{GeV}^{2}$, several curves show visible point-to-point variations because the numerical integral is very sensitive to the narrow region near $q_{ai}^2=0$. These variations are not physical resonances, and the curves become smoother when a larger cutoff removes this region.

Fig.~\ref{fig:LN-total} also shows that $nn\to pp e^-e^-$ is less sensitive to $\Delta q^2$ than the five hyperon channels. The difference is most visible at small $\Delta q^2$,
where the region near $q_{ai}^2=0$ makes a larger relative contribution to the hyperon cross sections.
For each cutoff value, Fig.~\ref{fig:LN-total} shows that the $nn$ channel has the largest cross section in the energy region where all six channels are allowed. The $n\Lambda$ channel is generally the
second largest, while $\Lambda\Sigma^0$ and $\Sigma^0\Sigma^0$ are the most suppressed.  The cross sections in Tabs.~\ref{tab:NN_Dq2_combined}--\ref{tab:SS_Dq2_combined} are compared below, together with a discussion of how the CKM factors affect their ordering.

\begin{figure}[t]
\begin{center}
\includegraphics[scale=0.88]{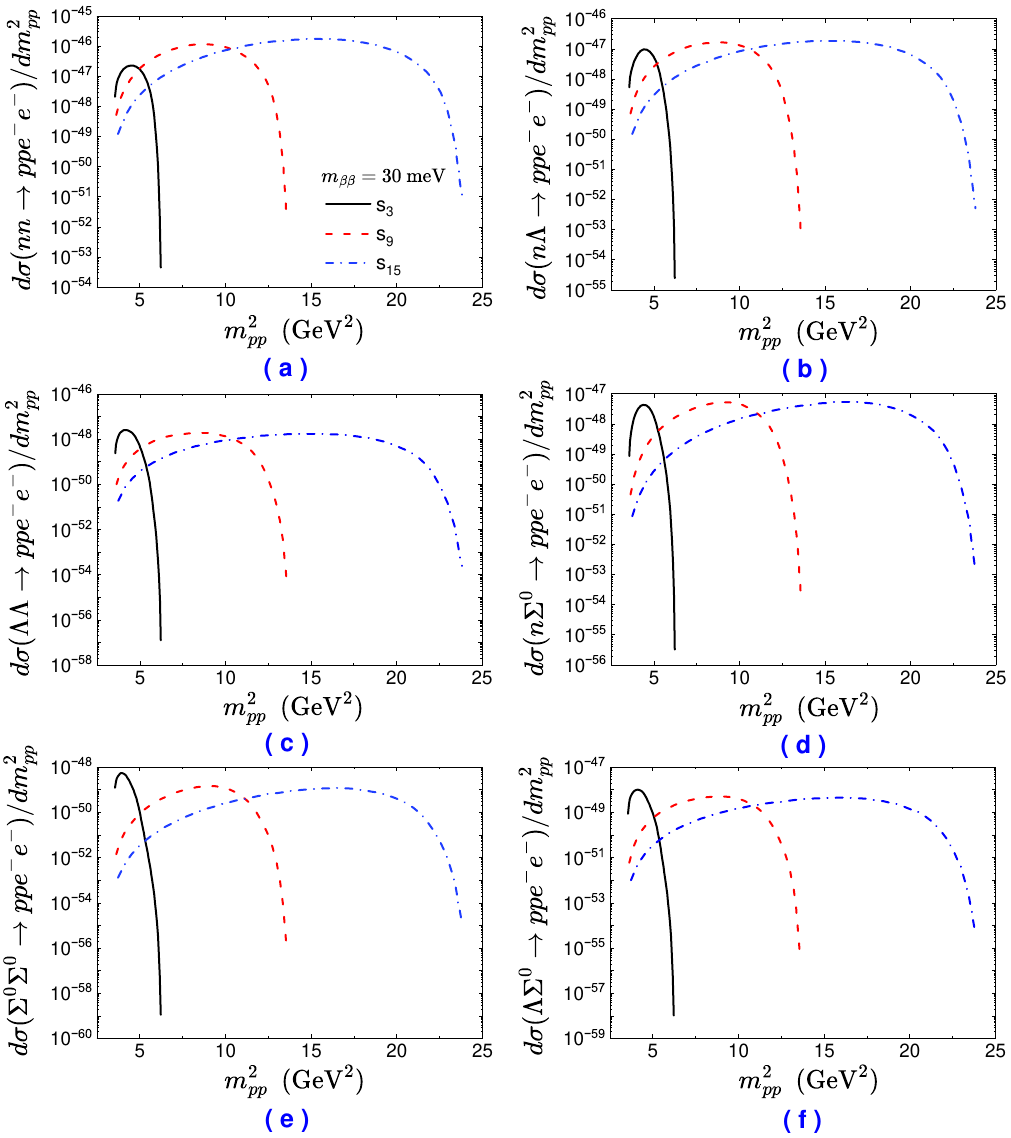}
\end{center}
\caption{Differential cross sections $d\sigma/dm_{pp}^2$ for the light-Majorana-neutrino contribution with $m_{\beta\beta}=30~\mathrm{meV}$ at three representative CM energies,
$\sqrt{s}=2.5$, $3.7$, and $4.9~\mathrm{GeV}$ (denoted by $s_3$, $s_9$, and $s_{15}$ in the figure). Panels (a)--(f) correspond to $nn$, $n\Lambda$, $\Lambda\Lambda$, $n\Sigma^0$, $\Sigma^0\Sigma^0$,
and $\Lambda\Sigma^0\to pp e^-e^-$, respectively. The spacelike off-shell prescription $q_{ai}^2<-\Delta q^2$ with $\Delta q^2=10^{-2}~\mathrm{GeV}^2$ is imposed as described in Sec.~\ref{sec:framework}. The differential cross sections are given in
$\mathrm{GeV}^{-4}$.}
\label{fig:LN-mpp2}
\end{figure}

The invariant-mass spectra provide information beyond the total cross sections. Fig.~\ref{fig:LN-mpp2} shows a wide peak in each $m_{pp}^2$ spectrum. The peak position and spectral shape reflect the four-body phase space, baryonic transition form factors, and light-neutrino propagators. For $\sqrt{s}=2.5$, $3.7$, and $4.9~\mathrm{GeV}$, the kinematic upper limits of $m_{pp}^2$ are about $6.25$, $13.7$, and $24.0~\mathrm{GeV}^2$, respectively. As $\sqrt{s}$ increases, the peak moves to larger $m_{pp}^2$. Toward the upper end of the plotted range, each spectrum decreases as the available phase space becomes smaller.

\begin{figure}[t]
\begin{center}
\includegraphics[scale=0.88]{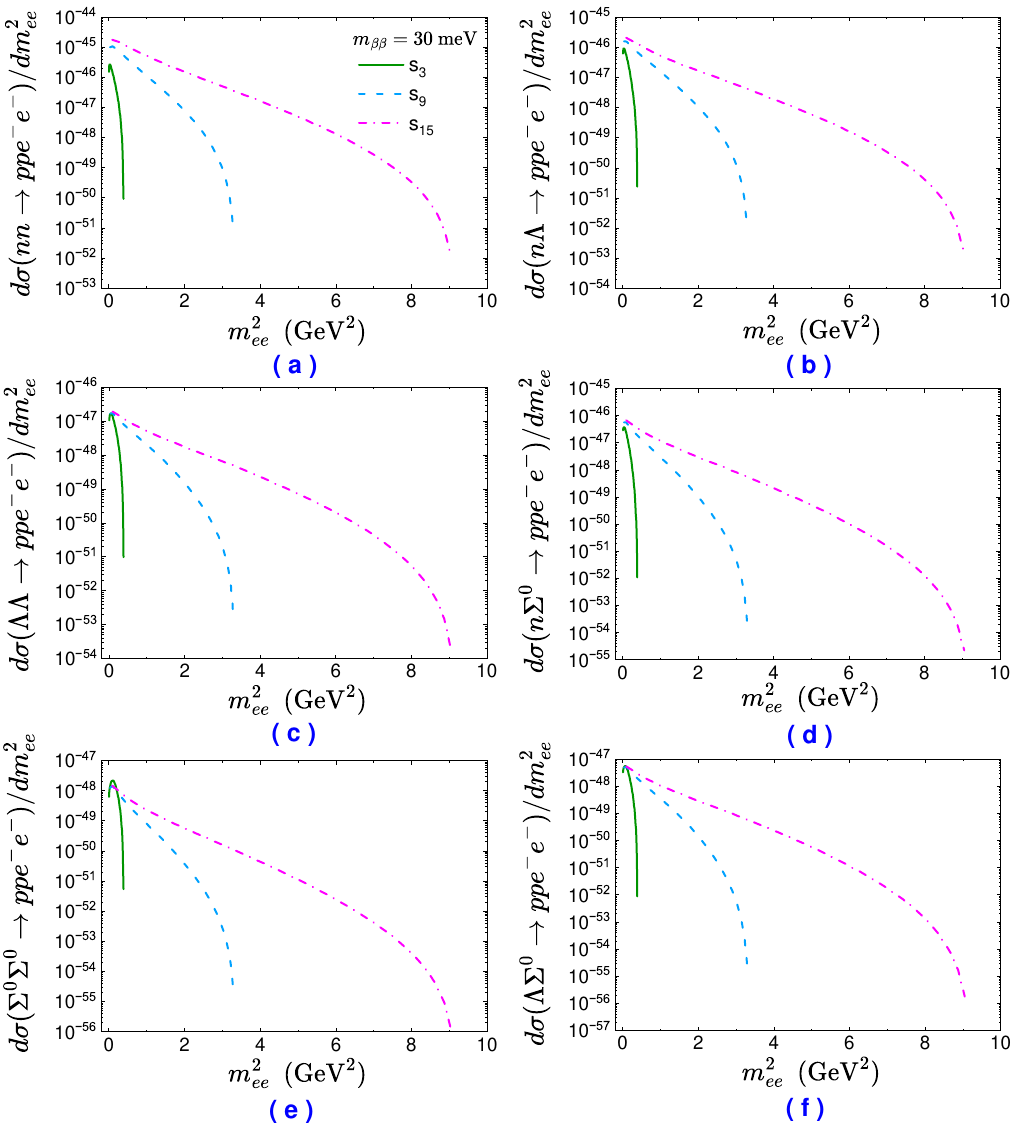}
\end{center}
\caption{Differential cross sections $d\sigma/dm_{ee}^2$ for the light-Majorana-neutrino contribution with $m_{\beta\beta}=30~\mathrm{meV}$ at $\sqrt{s}=2.5$, $3.7$, and $4.9~\mathrm{GeV}$
(denoted by $s_3$, $s_9$, and $s_{15}$ in the figure). Panels (a)-(f) correspond to $nn$, $n\Lambda$, $\Lambda\Lambda$, $n\Sigma^0$, $\Sigma^0\Sigma^0$, and $\Lambda\Sigma^0\to pp e^-e^-$, respectively.
The spacelike off-shell prescription $q_{ai}^2<-\Delta q^2$ with $\Delta q^2=10^{-2}~\mathrm{GeV}^2$ is imposed as described in Sec.~\ref{sec:framework}. The differential cross sections are given in
$\mathrm{GeV}^{-4}$.}
\label{fig:LN-mee2}
\end{figure}

In contrast, Fig.~\ref{fig:LN-mee2} shows that the $m_{ee}^2$ spectra are concentrated at small $m_{ee}^2$ and decrease toward the upper end of the plotted range. For $\sqrt{s}=2.5$, $3.7$, and $4.9~\mathrm{GeV}$, the kinematic upper limits of $m_{ee}^2$, given by $(\sqrt{s}-2m_p)^2$, are about $0.39$, $3.33$, and $9.14~\mathrm{GeV}^2$, respectively. Over these three energies, increasing $\sqrt{s}$ extends the allowed $m_{ee}^2$ range and gives a larger total cross section. The six channels have generally similar spectral shapes, while their overall sizes follow the ordering of the total cross sections.

Since the light-neutrino amplitude is proportional to $m_{\beta\beta}$, the cross section obeys
\begin{equation}
\frac{\sigma\left(m_{\beta\beta}\right)}{\sigma\left(m_{\beta\beta,0}\right)}
=\left(\frac{m_{\beta\beta}}{m_{\beta\beta,0}}\right)^2,
\label{eq:mbb-scaling}
\end{equation}
where $m_{\beta\beta,0}$ is any chosen value of the effective Majorana mass.
At fixed $\sqrt{s}$ and $\Delta q^2$, the results for $m_{\beta\beta}=10$, $30$, $50$, and $100~\mathrm{meV}$ are $100$, $900$, $2500$, and $10^4$ times the $1~\mathrm{meV}$ results, respectively.
Both the values at $\Delta q^2=10^{-2}~\mathrm{GeV}^2$ and their cutoff variations follow this quadratic scaling.

For a direct comparison of the six channels, the cross sections at $m_{\beta\beta}=30~\mathrm{meV}$, $\sqrt{s}=5.1~\mathrm{GeV}$, and $\Delta q^2=1.0\times10^{-2}~\mathrm{GeV}^2$ are
\begin{equation}
\begin{aligned}
\sigma_{nn}&=1.78\times10^{-45}~\mathrm{GeV}^{-2},\\
\sigma_{n\Lambda}&=1.84\times10^{-46}~\mathrm{GeV}^{-2},\\
\sigma_{n\Sigma^0}&=4.64\times10^{-47}~\mathrm{GeV}^{-2},\\
\sigma_{\Lambda\Lambda}&=1.72\times10^{-47}~\mathrm{GeV}^{-2},\\
\sigma_{\Lambda\Sigma^0}&=3.97\times10^{-48}~\mathrm{GeV}^{-2},\\
\sigma_{\Sigma^0\Sigma^0}&=9.28\times10^{-49}~\mathrm{GeV}^{-2}.
\end{aligned}
\label{eq:light-cross-sections}
\end{equation}
These results give the following hierarchy at the selected energy
\begin{equation}
\sigma_{nn}>\sigma_{n\Lambda}>\sigma_{n\Sigma^0}>\sigma_{\Lambda\Lambda}>\sigma_{\Lambda\Sigma^0}>\sigma_{\Sigma^0\Sigma^0}.
\label{eq:RH}
\end{equation}
The cross sections span more than three orders of magnitude, with the $nn$ result about $1.9\times10^3$ times larger than the $\Sigma^0\Sigma^0$ result.
This hierarchy is mainly explained by the CKM factors at the two charged-current vertices. The $nn$ amplitude contains $V_{ud}^2$, while amplitudes with one or two
strangeness-changing transitions contain $V_{ud}V_{us}$ or $V_{us}^2$, respectively. The corresponding cross sections are therefore proportional to $|V_{ud}|^4$,
$|V_{ud}V_{us}|^2$, and $|V_{us}|^4$. Since $|V_{us}|/|V_{ud}|\simeq0.23$, the strangeness-changing channels are suppressed. Within each group with the same CKM factor, the remaining
differences come from their baryonic transition form factors, initial-state masses, and phase space.

To quantify the cutoff dependence, $\Delta q^2=1.0\times10^{-2}~\mathrm{GeV}^2$ is used as the comparison value and is varied to $3.0\times10^{-3}$ and
$3.0\times10^{-2}~\mathrm{GeV}^2$.  Denoting the cross section evaluated at $\Delta q^2=x~\mathrm{GeV}^2$ by $\sigma_x$, the variations are defined by
\begin{equation}
\begin{aligned}
\sigma_{\rm c} &\equiv \sigma_{0.01}, \\
\Delta\sigma_+ &\equiv \sigma_{0.003}-\sigma_{\rm c},& \delta_+ &\equiv\frac{\Delta\sigma_+}{\sigma_{\rm c}}\times100\%, \\
\Delta\sigma_- &\equiv \sigma_{\rm c}-\sigma_{0.03},& \delta_- &\equiv-\frac{\Delta\sigma_-}{\sigma_{\rm c}}\times100\%.
\end{aligned}
\label{eq:cutoff-variation}
\end{equation}
Here, $\Delta\sigma_+$ is the increase in the cross section when the cutoff is lowered from $1.0\times10^{-2}$ to $3.0\times10^{-3}~\mathrm{GeV}^2$, while $\Delta\sigma_-$ is the
magnitude of the decrease when the cutoff is raised to $3.0\times10^{-2}~\mathrm{GeV}^2$. The quantities $\delta_+$ and $\delta_-$ are the corresponding percentage changes relative to
$\sigma_{\rm c}$. Thus, the positive percentages quoted below are values of $\delta_+$, while the negative percentages are values of $\delta_-$. The results are therefore written as
$\left(\sigma_{\rm c}\right)^{+\Delta\sigma_+}_{-\Delta\sigma_-}$.

Tabs.~\ref{tab:NN_Dq2_combined}--\ref{tab:SS_Dq2_combined} in Appendix~\ref{AppendixB} list these cross sections in $\mathrm{GeV}^{-2}$
for all six channels and show their dependence on $m_{\beta\beta}$, $\sqrt{s}$, and $\Delta q^2$. They confirm that $\sigma_{0.003}>\sigma_{0.01}>\sigma_{0.03}$.
At the smallest CM energy listed for each channel in Tabs.~\ref{tab:NN_Dq2_combined}--\ref{tab:SS_Dq2_combined}, the relative upper variations range from about $+69\%$ to $+357\%$, while the lower variations range from about $-65\%$ to $-95\%$. These energies lie closest to the corresponding channel thresholds and show the strong sensitivity to the propagator boundary in this region.
At $\sqrt{s}=5.1~\mathrm{GeV}$, the cutoff dependence is weaker than near threshold. The unrounded results underlying Tab.~\ref{tab:NN_Dq2_combined} give
$^{+3.1\%}_{-11.7\%}$ for the $nn$ channel. The corresponding rows in Tabs.~\ref{tab:LL_Dq2_combined}--\ref{tab:SS_Dq2_combined} give upper
variations of about $+31\%$ to $+68\%$ and lower variations of about $-23\%$ to $-33\%$ for the five hyperon channels. Thus, the $nn$ result becomes more
stable away from threshold, while the hyperon channels remain more sensitive to the propagator-enhanced region.
The increases and decreases given above result only from changing the off-shell cutoff. They are not statistical uncertainties and do not include uncertainties from the baryon transition form factors or the
hadronic-current approximation.

%\newpage

\subsection{Heavy-neutrino exchange}

For heavy-Majorana-neutrino exchange with $m_N\gtrsim50~\mathrm{GeV}$, the leading propagator correction is below
one percent over the phase-space region considered. In this limit, the neutrino mass and mixing enter only through $\eta_N$. Both the total and differential cross
sections therefore satisfy $\sigma(\eta_N)\propto\eta_N^2$.  The parameter values $\eta_N^A=4.0\times10^{-9}~\mathrm{GeV}^{-1}$ and
$\eta_N^B=8.0\times10^{-9}~\mathrm{GeV}^{-1}$ are used.  Therefore,
\begin{equation}
\frac{\sigma(\eta_N^B)}{\sigma(\eta_N^A)}=\left(\frac{\eta_N^B}{\eta_N^A}\right)^2=4.
\label{eq:eta-ratio}
\end{equation}
This relation holds at every energy and in every invariant-mass bin. The two parameter values therefore give the same dependence on $\sqrt{s}$ and the same spectral shapes, differing only in normalization.

\begin{figure}[t]
   \begin{center}
\includegraphics[scale=0.88]{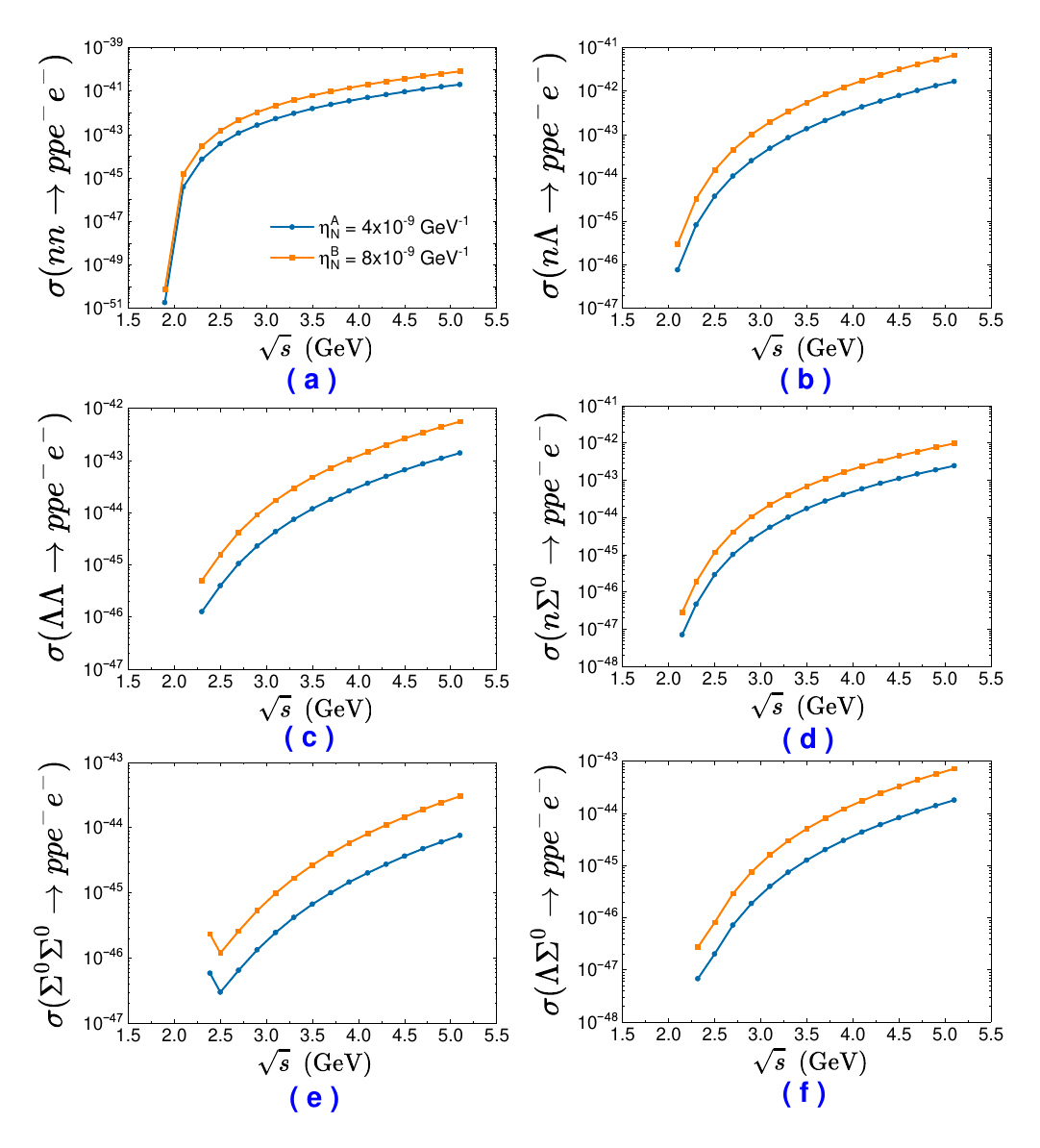}
\end{center}
\caption{Total cross sections for the heavy-Majorana-neutrino contribution as functions of $\sqrt{s}$ for the parameter values $\eta_N^A=4.0\times10^{-9}~\mathrm{GeV}^{-1}$ and $\eta_N^B=8.0\times10^{-9}~\mathrm{GeV}^{-1}$. Since $\sigma\propto\eta_N^2$ in the heavy-neutrino limit, the $\eta_N^B$ result is four times the corresponding $\eta_N^A$ result. The cross sections are given in $\mathrm{GeV}^{-2}$.}
\label{fig:HN-total}
\end{figure}

Fig.~\ref{fig:HN-total} shows the cross sections for heavy-Majorana-neutrino exchange as functions of $\sqrt{s}$. The $nn$ cross section rises rapidly just above threshold, while the $n\Lambda$, $\Lambda\Lambda$, $n\Sigma^0$, and $\Lambda\Sigma^0$ cross sections increase smoothly. The $\Sigma^0\Sigma^0$ cross section is the only one that first decreases, reaches a minimum near $\sqrt{s}=2.5~\mathrm{GeV}$, and then increases. Compared with the light-neutrino results calculated with $\Delta q^2=10^{-2}~\mathrm{GeV}^2$, the clearest differences occur near threshold, where several hyperon cross sections remain nearly flat or decrease before increasing. At each energy considered with $\sqrt{s}\geq2.9~\mathrm{GeV}$, the two exchange cases have the same channel ordering, with the $nn$ channel largest. Over this energy range, all six cross sections increase with $\sqrt{s}$.

\begin{figure*}[t]
   \begin{center}
\includegraphics[scale=0.88]{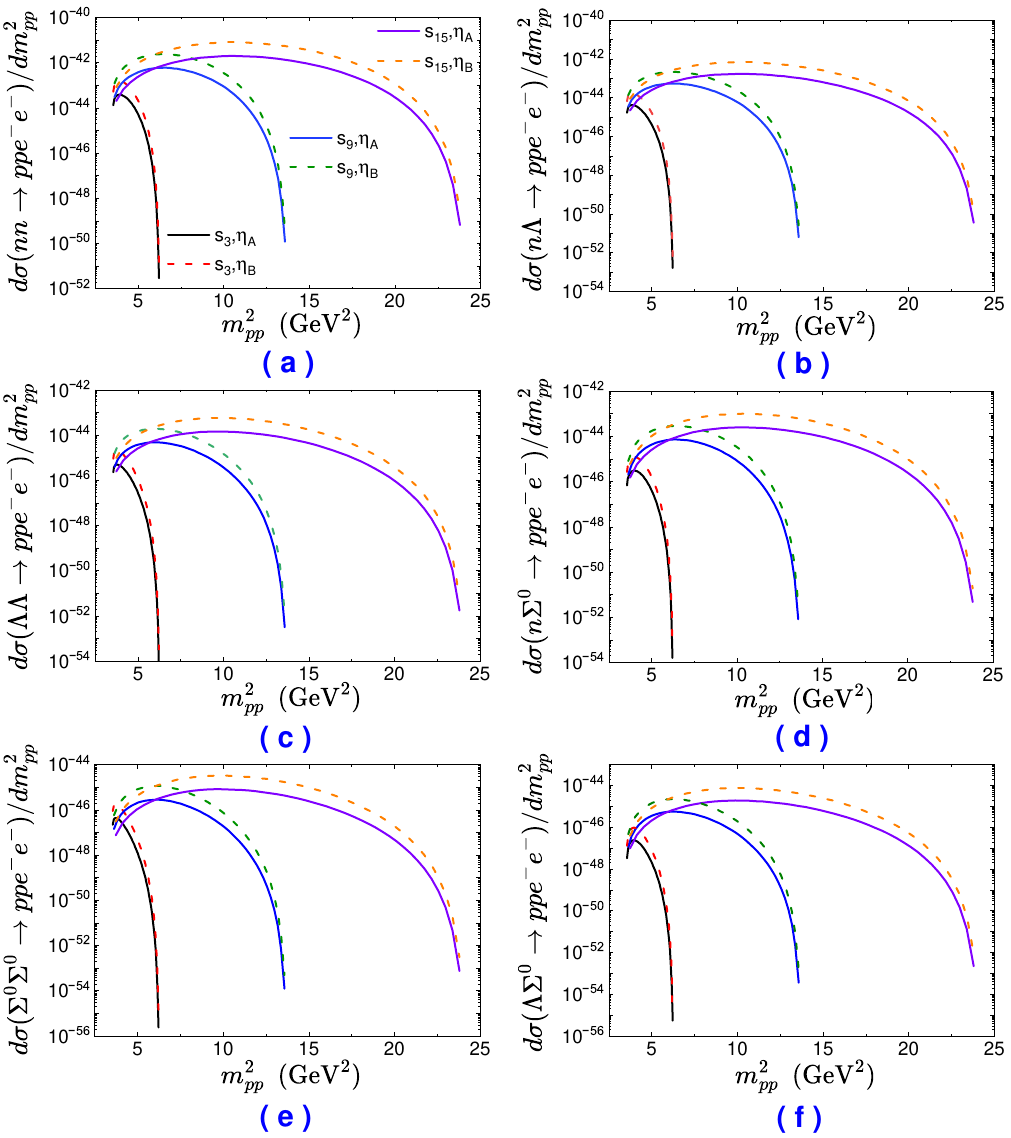}
\end{center}
\caption{Differential cross sections $d\sigma/dm_{pp}^2$ for the heavy-Majorana-neutrino contribution at $\sqrt{s}=2.5$, $3.7$, and $4.9~\mathrm{GeV}$ (denoted by $s_3$, $s_9$, and $s_{15}$), for $\eta_N^A=4.0\times10^{-9}~\mathrm{GeV}^{-1}$ and $\eta_N^B=8.0\times10^{-9}~\mathrm{GeV}^{-1}$. For each energy, the two parameter distributions differ by a factor of four. The differential cross sections are given in
$\mathrm{GeV}^{-4}$.}
\label{fig:HN-mpp2}
\end{figure*}

The invariant-mass distributions provide more detailed information about the four-body final state than the total cross sections. Fig.~\ref{fig:HN-mpp2} shows the $m_{pp}^2$ distributions for heavy-Majorana-neutrino exchange, with a peak in each spectrum above the two-proton threshold. At fixed $\sqrt{s}$, these peaks occur at smaller $m_{pp}^2$ than those in the light-neutrino spectra calculated with $\Delta q^2=10^{-2}~\mathrm{GeV}^2$ in Fig.~\ref{fig:LN-mpp2}. At $\sqrt{s}=3.7$ and $4.9~\mathrm{GeV}$, the light- and heavy-neutrino spectra have clearly different peak positions. For the $nn$ channel at $\sqrt{s}=4.9~\mathrm{GeV}$, Figs.~\ref{fig:LN-mpp2}(a) and \ref{fig:HN-mpp2}(a) show that the light-neutrino spectrum peaks at $m_{pp}^2\simeq15~\mathrm{GeV}^2$, while the heavy-neutrino spectrum peaks at $m_{pp}^2\simeq11~\mathrm{GeV}^2$. For both exchange cases, the peaks move to larger $m_{pp}^2$ as $\sqrt{s}$ increases.

\begin{figure*}[t]
\begin{center}
\includegraphics[scale=0.88]{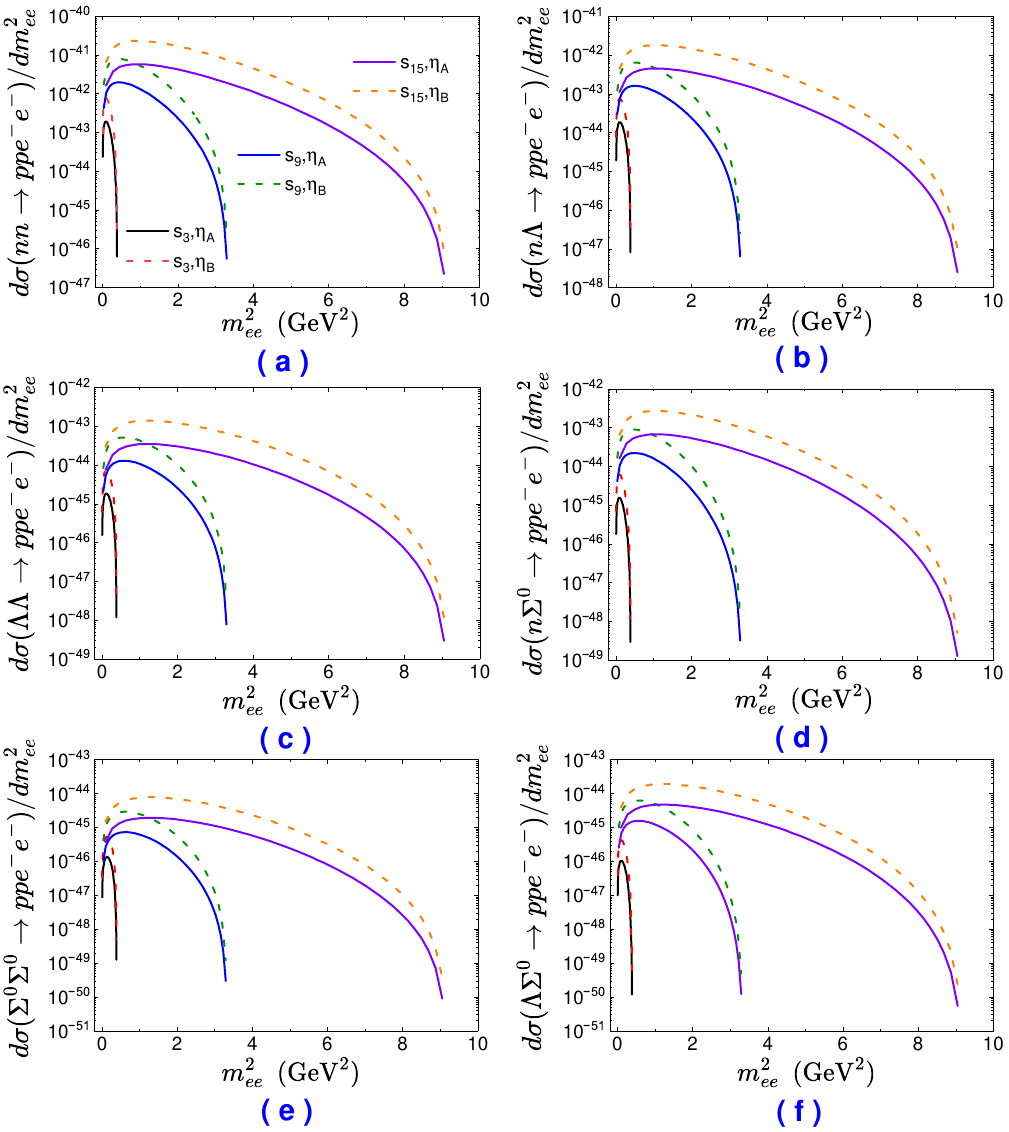}
\end{center}
\caption{Differential cross sections $d\sigma/dm_{ee}^2$ for the heavy-Majorana-neutrino contribution at $\sqrt{s}=2.5$, $3.7$, and $4.9~\mathrm{GeV}$ (denoted by $s_3$, $s_9$, and $s_{15}$), for $\eta_N^A=4.0\times10^{-9}~\mathrm{GeV}^{-1}$ and $\eta_N^B=8.0\times10^{-9}~\mathrm{GeV}^{-1}$. For each energy, the two parameter distributions differ by a factor of four. The differential cross sections are given in
$\mathrm{GeV}^{-4}$.}
 \label{fig:HN-mee2}
\end{figure*}

Fig.~\ref{fig:HN-mee2} shows the $m_{ee}^2$ distributions for heavy-Majorana-neutrino exchange. Like the light-neutrino spectra in Fig.~\ref{fig:LN-mee2}, they are concentrated at small $m_{ee}^2$. At fixed $\sqrt{s}$, however, the heavy-neutrino spectra peak at larger $m_{ee}^2$ than the light-neutrino spectra calculated with $\Delta q^2=10^{-2}~\mathrm{GeV}^2$, whose maxima lie closer to the dielectron threshold. For the $nn$ channel at $\sqrt{s}=4.9~\mathrm{GeV}$, Fig.~\ref{fig:HN-mee2}(a) shows that the heavy-neutrino spectrum peaks near $m_{ee}^2=1~\mathrm{GeV}^2$.

\begin{figure*}[t]
\begin{center}
\includegraphics[scale=0.88]{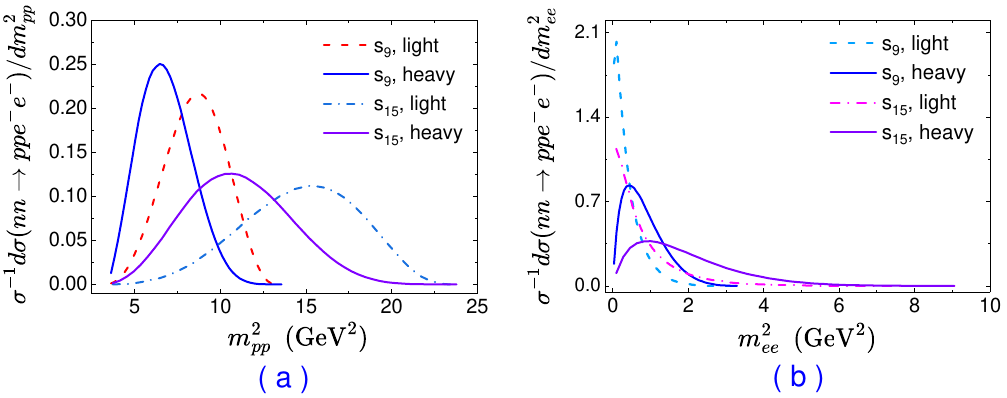}
\end{center}
\caption{Normalized invariant-mass spectra for
$nn\to pp e^-e^-$. Panels (a) and (b) show
$\sigma^{-1}d\sigma/dm_{pp}^2$ and
$\sigma^{-1}d\sigma/dm_{ee}^2$, respectively, at
$\sqrt{s}=3.7$ and $4.9~\mathrm{GeV}$. The light-neutrino
results are calculated with $m_{\beta\beta}=30~\mathrm{meV}$
and $\Delta q^2=10^{-2}~\mathrm{GeV}^2$, while the
heavy-neutrino results use $\eta_N^A=4.0\times10^{-9}
~\mathrm{GeV}^{-1}$. Each spectrum is normalized to unit
area. The normalized distributions are in
$\mathrm{GeV}^{-2}$.}
\label{fig:normalized-spectra}
\end{figure*}

For fixed $\sqrt{s}$, the allowed range of $m_{pp}^2$ is the same for the two exchange cases, and so is that of $m_{ee}^2$. These ranges are determined only by the external-particle masses and $\sqrt{s}$. To compare the
spectral shapes directly, each distribution is divided by its own total cross section. Fig.~\ref{fig:normalized-spectra} shows the normalized $nn$ spectra at $\sqrt{s}=3.7$ and $4.9~\mathrm{GeV}$.  After this normalization, the heavy-neutrino $m_{pp}^2$ spectrum still peaks at a smaller value than the light-neutrino spectrum. The heavy-neutrino $m_{ee}^2$ spectrum still peaks at a larger value than the light-neutrino spectrum.
The shifts in the peak positions mainly reflect the different momentum dependence of the exchanged-neutrino propagator. In the light-neutrino limit, the propagator contributes a factor $1/q_{ai}^2$, whereas the leading term in the heavy-neutrino limit is independent of $q_{ai}^2$. The normalized spectra therefore provide information beyond the overall cross-section normalization. The peak differences shown here depend on the light-neutrino cutoff
and the adopted baryon form factors. Normalization to unit area does not remove this dependence.
 More generally, differential distributions have been studied as a way to distinguish different mechanisms of $0\nu\beta\beta$ decay \cite{Graf:2022Mechanisms}.


Tabs.~\ref{tab:HN_NN_NL}--\ref{tab:HN_SL_SS} in Appendix~\ref{AppendixB} list the cross sections for all six channels at the two heavy-neutrino parameter values. For $\eta_N^A$, increasing $\sqrt{s}$ from $2.5$ to $5.1~\mathrm{GeV}$ increases the cross sections by factors ranging from about $2.5\times10^2$ to $8.9\times10^2$. These factors describe the energy dependence within each channel. Since the results for $\eta_N^B$ are four times those for $\eta_N^A$ at every energy, the two parameter values give the same energy dependence and channel ordering. For a direct comparison of the six channels at fixed energy, the results for $\eta_N^A$ at $\sqrt{s}=5.1~\mathrm{GeV}$ are
\begin{equation}
\begin{aligned}
\sigma_{nn}&=1.98\times10^{-41}~\mathrm{GeV}^{-2},\\
\sigma_{n\Lambda}&=1.66\times10^{-42}~\mathrm{GeV}^{-2},\\
\sigma_{n\Sigma^0}&=2.42\times10^{-43}~\mathrm{GeV}^{-2},\\
\sigma_{\Lambda\Lambda}&=1.39\times10^{-43}~\mathrm{GeV}^{-2},\\
\sigma_{\Lambda\Sigma^0}&=1.79\times10^{-44}~\mathrm{GeV}^{-2},\\
\sigma_{\Sigma^0\Sigma^0}&=7.55\times10^{-45}~\mathrm{GeV}^{-2}.
\end{aligned}
\label{eq:heavy-cross-sections}
\end{equation}
These values follow the same ordering as the light-neutrino results in Eq.~\eqref{eq:RH}. The $nn$ cross section is about $2.6\times10^3$ times the $\Sigma^0\Sigma^0$ value.

At lower energies, however, two pairs of channels change their relative order. Tab.~\ref{tab:HN_NS_LL} gives $\sigma_{\Lambda\Lambda}>\sigma_{n\Sigma^0}$ at $\sqrt{s}=2.5$ and $2.7~\mathrm{GeV}$, whereas $\sigma_{n\Sigma^0}>\sigma_{\Lambda\Lambda}$ at all energies considered with $\sqrt{s}\geq2.9~\mathrm{GeV}$. Tab.~\ref{tab:HN_SL_SS} gives $\sigma_{\Sigma^0\Sigma^0}>\sigma_{\Lambda\Sigma^0}$ at $\sqrt{s}=2.5~\mathrm{GeV}$, whereas $\sigma_{\Lambda\Sigma^0}>\sigma_{\Sigma^0\Sigma^0}$ at all energies considered with $\sqrt{s}\geq2.7~\mathrm{GeV}$. Thus, among the energies considered, the hierarchy in Eq.~\eqref{eq:RH} holds for $\sqrt{s}\geq2.9~\mathrm{GeV}$.

%\clearpage
\section{Summary}
\label{sec:summary}

Lepton-number violation is closely connected to the origin of neutrino mass and the possible Majorana nature of neutrinos. The free-baryon reactions $B_1B_2\to pp e^-e^-$ were studied for six nonstrange and strange initial states, extending the charged-current transitions from $d\to u$ to $s\to u$. Light- and heavy-Majorana-neutrino exchange were treated in the same baryon-level framework, including momentum-dependent weak form factors, all exchange terms associated with the identical final-state fermions, and exact four-body phase space.

For light-neutrino exchange, the cross sections scale with the square of the effective Majorana mass. The calculated cross sections increase as the off-shell cutoff $\Delta q^2$ is reduced because more of the spacelike phase space near $q_{ai}^2=0$ is retained.
 This cutoff dependence is strongest near threshold and in the hyperon channels, whereas the $nn$ result becomes less sensitive at higher energies. The quoted variations describe only the cutoff dependence and are not statistical uncertainties.

For heavy-neutrino exchange, the neutrino mass and mixing enter through $\eta_N=|U_{eN}|^2/m_N$, and the cross sections scale with $\eta_N^2$. The leading propagator correction remains below one percent for $m_N\gtrsim50~\mathrm{GeV}$ over the phase-space region considered. No spacelike cutoff is needed in this heavy-neutrino-mass regime, and changing $\eta_N$ only rescales the total and differential cross sections.

For the calculated energies $\sqrt{s}\geq2.9~\mathrm{GeV}$, the two exchange cases give the same channel ordering, with the $nn$ cross section largest mainly because of the CKM factors. Baryon masses,
transition form factors, and phase space produce further differences between channels, while the ordering changes near threshold for heavy-neutrino exchange. The invariant-mass distributions provide an
additional difference between the two exchange cases. At fixed $\sqrt{s}$, compared with the normalized light-neutrino spectra calculated with $\Delta q^2=10^{-2}~\mathrm{GeV}^2$, the peak of the
normalized heavy-neutrino $m_{pp}^2$ spectrum occurs at a smaller value, whereas the peak of the normalized heavy-neutrino $m_{ee}^2$ spectrum occurs at a larger value.
 Thus, the comparison between the two exchange cases is not limited to the overall sizes of the cross sections.

Experimentally, BESIII provides large hyperon samples for studies of baryon interactions and rare baryon processes \cite{BESIIIHyperonReview:2026}, although the predicted cross sections are far below its current sensitivity. The proposed Super Tau-Charm Facility~\cite{Achasov:2023gey}, with much higher luminosity and substantially larger hyperon yields, could improve information on hyperon production and baryon--baryon interactions. In principle, intense neutron beams and dedicated neutron-correlation measurements offer a novel way to explore the free-baryon analogue of $0\nu\beta\beta$ decay. Nevertheless, these results provide theoretical reference calculations for lepton-number-violating free-baryon reactions and a framework that can be extended to other sources of lepton-number violation.

\section*{ACKNOWLEDGEMENTS}
The work was supported by the National Natural Science Foundation of China (Nos. 12665018, 12365014 and 12225509).

%\newpage
\begin{appendix}
 \begin{center}
 {\LARGE{\bf Appendix}}
 \end{center}

\section{\hspace{-0.6cm} Details of the theoretical framework} \label{AppendixA}

\subsection{The form factors}
The vector and axial-vector form factors, $g_V^{Bp}$ and $g_A^{Bp}$, are given here.  Their momentum dependence is described by the dipole form
\begin{equation}
g_{V,A}^{Bp}(t)=\frac{g_{V,A}^{Bp}(0)}{\left(1-t/M_{V,A}^2\right)^2},
\label{eq:dipole}
\end{equation}
where the parameters $M_V$ and $M_A$ are the vector and axial-vector dipole masses.
Commonly used parametrizations of neutron and hyperon semileptonic transitions are adopted \cite{Simkovic:1999,Cabibbo:2003,Ratcliffe:2004jt,Mateu:2005wi}.
Tab.~\ref{tab:FFinput} lists the vector and axial-vector form-factor parameters used in this work \cite{Cabibbo:2003}. Here, $F$ and $D$ are the two axial couplings in the flavor-SU(3) description of baryon transitions, and the dipole masses are given in GeV.
\begin{table}[h]
\centering
\caption{Vector and axial-vector form-factor parameters, and  $F+\frac{D}{3}=0.731$ as well as $D-F=0.341$.}
\label{tab:FFinput}
 \renewcommand{\arraystretch}{1.5}
 \setlength{\tabcolsep}{20pt}
\begin{tabular}{ccccc}\hline\hline
Transition          & $g_V(0)$       & $g_A(0)$                &$M_V$      & $M_A$\\\hline
$n\to p$            &$1$             &$1.267$                  &$0.84$     &$1.08$\\\hline
$\Lambda\to p$      &$-\sqrt{\frac{3}{2}}$   &$-\sqrt{\frac{3}{2}}(F+\frac{D}{3})$            &$0.97$     &$1.25$\\\hline
$\Sigma^0\to p$     &$-\sqrt{\frac{1}{2}}$   &$\sqrt{\frac{1}{2}}(D-F)$             &$0.97$     &$1.25$\\\hline
\end{tabular}
\end{table}

The dipole form is used as a common parametrization over the
momentum-transfer range considered. No numerical uncertainty is
assigned to this choice, especially for the hyperon transitions, so
the absolute cross sections depend on the adopted baryon form
factors. A quantitative uncertainty estimate would require more
precise momentum-dependent form-factor input.

\subsection{An example of \texorpdfstring{$T_{mn}$}{Tmn}}
As an example, the diagonal term $T_{11}$ associated with $q_{31}$ can be written as
\begin{align}
T_{11}= 8\left|\sum_{i'}\frac{U_{ei'}^2m_{\nu_{i'}}}{\left(q_{31}^2-m_{\nu_{i'}}^2+i0^+\right) }\right|^2 L_{12;\mu\nu\mu'\nu'}\,H_{31}^{\mu\mu'}\,H_{42}^{\nu\nu'} ,
\label{eq:T11}
\end{align}
with
\begin{align}
H_{31}^{\mu\mu'}={}&\mathrm{Tr}\!\left[(\slashed p_3+m_p)\gamma^\mu(g_V(D_{31}^2)-g_A(D_{31}^2)\gamma_5)(\slashed p_1+m_{B_1})\gamma^{\mu'}(g_V(D_{31}^2)-g_A(D_{31}^2)\gamma_5)\right],\label{eq:H31}\\
H_{42}^{\nu\nu'}={}&\mathrm{Tr}\!\left[(\slashed p_4+m_p)\gamma^\nu(g_V^{\prime}(D_{42}^2)-g_A^{\prime}(D_{42}^2)\gamma_5)(\slashed p_2+m_{B_2})\gamma^{\nu'}(g_V^{\prime}(D_{42}^2)-g_A^{\prime}(D_{42}^2)\gamma_5)\right],\label{eq:H42}\\
L_{12;\mu\nu\mu'\nu'}={}&\mathrm{Tr}\!\left[\slashed k_1\gamma_\mu\gamma_\nu\slashed k_2\gamma_{\nu'}\gamma_{\mu'}(1-\gamma_5)\right].\label{eq:L12}
\end{align}
The electron mass is retained through $k_i^2=m_e^2$ and in the phase-space limits. After applying the left-handed weak-current projectors, the explicit electron-mass terms in the spin sums vanish, giving Eq.~\eqref{eq:L12}.
The absolute square of the sum in Eq.~\eqref{eq:T11} includes the interference between different neutrino mass eigenstates. In the light-neutrino limit used in the numerical calculation, the neutrino-dependent factor in $T_{11}$ reduces to $8m_{\beta\beta}^2/(q_{31}^2)^2$. Here, $D_{a1}^2\equiv(p_a-p_1)^2$ and $D_{b2}^2\equiv(p_b-p_2)^2$ are the squared momentum transfers at the two baryonic vertices.
The unprimed and primed form factors correspond to the charged-current vertices involving $B_1$ and $B_2$, respectively. All 16 terms $T_{mn}$ are retained. The Dirac algebra and Lorentz contractions are
evaluated with FeynCalc \cite{Mertig:1990an,Shtabovenko:2016sxi}.

\subsection{Kinematic analysis}
\label{subsec:kinematics}

The incoming momenta in the overall CM frame are
\begin{equation}
p_1=(E_{B_1},0,0,p_z),\qquad p_2=(E_{B_2},0,0,-p_z),
\label{eq:p12}
\end{equation}
where
\begin{equation}
E_{B_1}=\frac{s+m_{B_1}^2-m_{B_2}^2}{2\sqrt{s}},~~~~~~E_{B_2}=\frac{s+m_{B_2}^2-m_{B_1}^2}{2\sqrt{s}},~~~~~~~p_z=\frac{\sqrt{\lambda(s,m_{B_1}^2,m_{B_2}^2)}}{2\sqrt{s}},
\label{eq:initialKin}
\end{equation}
where $s=(p_1+p_2)^2$, the total CM energy is $\sqrt{s}=E_{B_1}+E_{B_2}$, and the kinematic factor $\lambda(x,y,z)=x^2+y^2+z^2-2xy-2xz-2yz$.

\begin{figure}[t]
\centering
\includegraphics[width=0.85\linewidth]{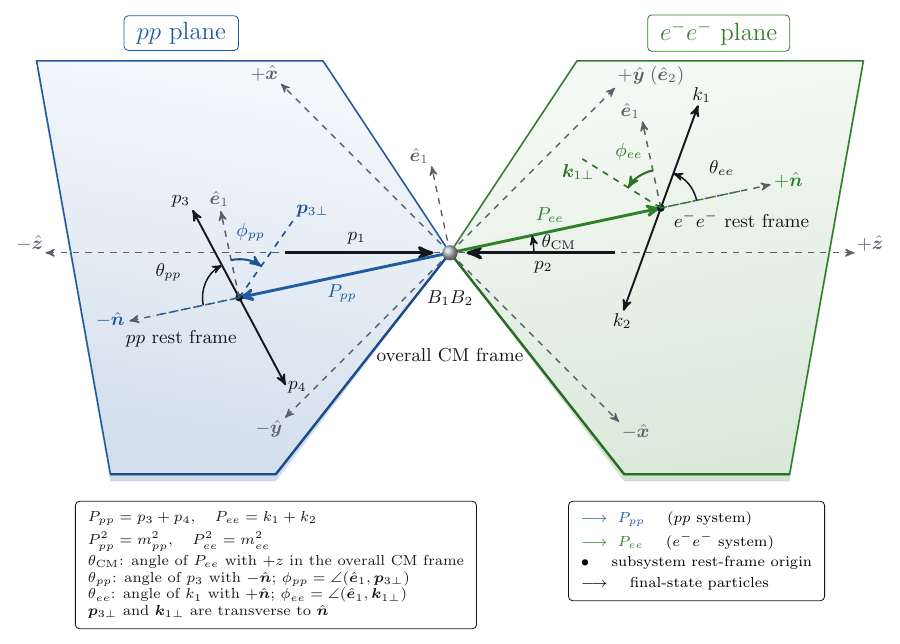}
\caption{Kinematic configuration for $B_1(p_1)B_2(p_2)\to p(p_3)p(p_4)e^-(k_1)e^-(k_2)$ in the overall CM frame.  The incoming baryons move along the
$\pm\hat{\bm z}$ directions.  The diproton and dielectron momenta, $P_{pp}$ and $P_{ee}$, are back-to-back along $-\hat{\bm n}$ and $+\hat{\bm n}$, respectively.
%The angle $\theta_{\rm CM}$ is measured between $P_{ee}$ and the $+\hat{\bm z}$ axis.  In the $pp$ and $e^-e^-$ rest frames, $(\theta_{pp},\phi_{pp})$ and $(\theta_{ee},\phi_{ee})$ specify the directions of $p_3$ and $k_1$, respectively.  The polar angles are measured from $-\hat{\bm n}$ and $+\hat{\bm n}$, while the azimuthal angles are measured from $\hat{\bm e}_1$ to the transverse projections $\bm p_{3\perp}$ and $\bm k_{1\perp}$.
The unit vectors $\hat{\bm e}_1$ and
$\hat{\bm e}_2$ define the transverse basis.  The drawing is
schematic.}
\label{fig:kinematic-configuration}
\end{figure}

The four-body final-state kinematics is parameterized in terms of the $pp$ and $e^-e^-$ subsystems by adapting the general two-subsystem construction of Ref.~\cite{Kim:2022nqq}.
 The coordinate system and the angular conventions used below are summarized in
Fig.~\ref{fig:kinematic-configuration}.
The total, diproton, and dielectron four-momenta are denoted by $P$, $P_{pp}$, and $P_{ee}$, respectively, with $P\equiv p_1+p_2=P_{pp}+P_{ee}$.
The subsystem momenta and invariant masses are defined by
\begin{equation}
P_{pp}\equiv p_3+p_4,\qquad P_{ee}\equiv k_1+k_2, \qquad P_{pp}^2=m_{pp}^2, \qquad P_{ee}^2=m_{ee}^2,
\label{eq:clusters}
\end{equation}
where  $m_{pp}$ and $m_{ee}$ are the invariant masses of the two subsystems. Their energies and common three-momentum magnitude are
\begin{equation}
E_{pp}=\frac{s+m_{pp}^2-m_{ee}^2}{2\sqrt{s}},\qquad E_{ee}=\frac{s-m_{pp}^2+m_{ee}^2}{2\sqrt{s}},\qquad Y=\frac{\sqrt{\lambda(s,m_{pp}^2,m_{ee}^2)}}{2\sqrt{s}} .
\label{eq:clusterKin}
\end{equation}
The direction of the dielectron subsystem in the overall CM frame is
written as
\begin{equation}
\hat{\bm n}=\left(\sin\theta_{\rm CM}\cos\phi_{\rm CM},\sin\theta_{\rm CM}\sin\phi_{\rm CM},\cos\theta_{\rm CM}\right),
\end{equation}
where $\theta_{\rm CM}$ is the polar angle of $P_{ee}$ with respect to
the $+z$ beam direction, and $\phi_{\rm CM}$ is its azimuthal angle
around the $z$ axis, measured from the $+x$ direction in the transverse
$x$--$y$ plane.  Since the unpolarized initial state is rotationally
symmetric about the beam axis, $\phi_{\rm CM}$ can be chosen as zero.
Consequently,
\begin{equation}
 \hat{\bm n}=(\sin\theta_{\rm CM},0,\cos\theta_{\rm CM}),
\end{equation}
and the integration over $\phi_{\rm CM}$ gives an overall factor of
$2\pi$.
The subsystem four-momenta are therefore
\begin{equation}
P_{ee}=(E_{ee},Y\hat{\bm n}),\qquad P_{pp}=(E_{pp},-Y\hat{\bm n}).
\label{eq:cluster4mom}
\end{equation}

In the $pp$ and $ee$ rest frames, the relative momentum magnitudes are
\begin{equation}
Y_{pp}=\frac{1}{2}\sqrt{m_{pp}^2-4m_p^2}, \qquad Y_{ee}=\frac{1}{2}\sqrt{m_{ee}^2-4m_e^2}.
\label{eq:YppYee}
\end{equation}
The transverse unit vectors are defined by
\begin{equation}
\hat{\bm  e}_1=(\cos\theta_{\rm CM},0,-\sin\theta_{\rm CM}),\qquad  \hat{\bm  e}_2=(0,1,0),
\end{equation}
The angles $(\theta_{pp},\phi_{pp})$ and $(\theta_{ee},\phi_{ee})$ specify the $p_3$ and $k_1$ directions in the $pp$ and $ee$ rest frames, respectively. The polar angles are measured from $-\hat{\bm n}$ and $+\hat{\bm n}$, respectively; both azimuthal angles are measured from $\hat{\bm e}_1$ toward $\hat{\bm e}_2$ in the transverse plane.
With these definitions, the final-state momenta can be expressed  as
\begin{align}
E_3&=\frac{E_{pp}}{2}+\frac{YY_{pp}}{m_{pp}}\cos\theta_{pp},  &  \vec{\bm p}_3&= Y_{pp}\sin\theta_{pp}(\cos\phi_{pp}\hat{\bm e}_1+\sin\phi_{pp}\hat{\bm e}_2)-\left(\frac{Y}{2}+\frac{E_{pp}Y_{pp}}{m_{pp}}\cos\theta_{pp}\right)\hat{\bm n},\\
E_4&=\frac{E_{pp}}{2}-\frac{YY_{pp}}{m_{pp}}\cos\theta_{pp},  &  \vec{\bm p}_4&=-Y_{pp}\sin\theta_{pp}(\cos\phi_{pp}\hat{\bm e}_1+\sin\phi_{pp}\hat{\bm e}_2)-\left(\frac{Y}{2}-\frac{E_{pp}Y_{pp}}{m_{pp}}\cos\theta_{pp}\right)\hat{\bm n},\\
E_{e_1}&=\frac{E_{ee}}{2}+\frac{YY_{ee}}{m_{ee}}\cos\theta_{ee},&\vec{\bm k}_1&=Y_{ee}\sin\theta_{ee}(\cos\phi_{ee}\hat{\bm e}_1+\sin\phi_{ee}\hat{\bm e}_2)+\left(\frac{Y}{2}+\frac{E_{ee}Y_{ee}}{m_{ee}}\cos\theta_{ee}\right)\hat{\bm n},\\
E_{e_2}&=\frac{E_{ee}}{2}-\frac{YY_{ee}}{m_{ee}}\cos\theta_{ee},&\vec{\bm k}_2&=-Y_{ee}\sin\theta_{ee}(\cos\phi_{ee}\hat{\bm e}_1+\sin\phi_{ee}\hat{\bm e}_2)+\left(\frac{Y}{2}-\frac{E_{ee}Y_{ee}}{m_{ee}}\cos\theta_{ee}\right)\hat{\bm n}.
\label{eq:momentaA}
\end{align}
The explicit four-vectors are
{\small \begin{eqnarray}
p_3 &=&
\begin{pmatrix}
\frac{E_{pp}}{2} + \frac{Y Y_{pp}}{m_{pp}} \cos\theta_{pp} \\[8pt]
Y_{pp} \sin\theta_{pp} \cos\phi_{pp} \cos\theta_{\text{CM}}
- \left( \frac{Y}{2} + \frac{E_{pp} Y_{pp}}{m_{pp}} \cos\theta_{pp} \right) \sin\theta_{\text{CM}} \\[8pt]
Y_{pp} \sin\theta_{pp} \sin\phi_{pp} \\[8pt]
- Y_{pp} \sin\theta_{pp} \cos\phi_{pp} \sin\theta_{\text{CM}}
- \left( \frac{Y}{2} + \frac{E_{pp} Y_{pp}}{m_{pp}} \cos\theta_{pp} \right) \cos\theta_{\text{CM}}
\end{pmatrix},\\[8pt]
%\end{equation*}
%\begin{equation*}
p_4 &=&
\begin{pmatrix}
\frac{E_{pp}}{2} - \frac{Y Y_{pp}}{m_{pp}} \cos\theta_{pp} \\[8pt]
- Y_{pp} \sin\theta_{pp} \cos\phi_{pp} \cos\theta_{\text{CM}}
+ \left( -\frac{Y}{2} + \frac{E_{pp} Y_{pp}}{m_{pp}} \cos\theta_{pp} \right) \sin\theta_{\text{CM}} \\[8pt]
- Y_{pp} \sin\theta_{pp} \sin\phi_{pp} \\[8pt]
Y_{pp} \sin\theta_{pp} \cos\phi_{pp} \sin\theta_{\text{CM}}
+ \left( -\frac{Y}{2} + \frac{E_{pp} Y_{pp}}{m_{pp}} \cos\theta_{pp} \right) \cos\theta_{\text{CM}}
\end{pmatrix},
\end{eqnarray}
\begin{eqnarray}
k_1&=&
\begin{pmatrix}
\frac{E_{ee}}{2} + \frac{Y Y_{ee}}{m_{ee}} \cos\theta_{ee} \\[8pt]
Y_{ee} \sin\theta_{ee} \cos\phi_{ee} \cos\theta_{\text{CM}}
+ \left( \frac{Y}{2} + \frac{E_{ee} Y_{ee}}{m_{ee}} \cos\theta_{ee} \right) \sin\theta_{\text{CM}} \\[8pt]
Y_{ee} \sin\theta_{ee} \sin\phi_{ee} \\[8pt]
- Y_{ee} \sin\theta_{ee} \cos\phi_{ee} \sin\theta_{\text{CM}}
+ \left( \frac{Y}{2} + \frac{E_{ee} Y_{ee}}{m_{ee}} \cos\theta_{ee} \right) \cos\theta_{\text{CM}}
\end{pmatrix},\\[8pt]
%\end{equation*}
%\begin{equation*}
k_2 &=&
\begin{pmatrix}
\frac{E_{ee}}{2} - \frac{Y Y_{ee}}{m_{ee}} \cos\theta_{ee} \\[8pt]
- Y_{ee} \sin\theta_{ee} \cos\phi_{ee} \cos\theta_{\text{CM}}
+ \left( \frac{Y}{2} - \frac{E_{ee} Y_{ee}}{m_{ee}} \cos\theta_{ee} \right) \sin\theta_{\text{CM}} \\[8pt]
- Y_{ee} \sin\theta_{ee} \sin\phi_{ee} \\[8pt]
Y_{ee} \sin\theta_{ee} \cos\phi_{ee} \sin\theta_{\text{CM}}
+ \left( \frac{Y}{2} - \frac{E_{ee} Y_{ee}}{m_{ee}} \cos\theta_{ee} \right) \cos\theta_{\text{CM}}
\end{pmatrix}.
\end{eqnarray}}
This parametrization satisfies the on-shell conditions for all external particles and conserves four-momentum, $p_1+p_2=p_3+p_4+k_1+k_2$.

\subsection{Four-body phase space}
\label{subsec:crosssection}

The final state is divided into diproton and dielectron subsystems.  The Lorentz-invariant four-body phase-space element $d\Phi_4$ can be factorized as \cite{Chen:2021FourLepton}
\begin{align}
d\Phi_4={}&\frac{dm_{pp}^2}{2\pi}\frac{dm_{ee}^2}{2\pi}d\Phi_2(P;P_{pp},P_{ee})d\Phi_2(P_{pp};p_3,p_4)d\Phi_2(P_{ee};k_1,k_2),\label{eq:dPhi4factor}
\end{align}
with
\begin{equation}
d\Phi_2(Q;a,b)=\frac{|\vec{\bm p}|}{16\pi^2\sqrt{Q^2}}\,d\Omega .
\end{equation}
Here, $d\Phi_2(Q;a,b)$ is the two-body phase-space element for a system with four-momentum $Q$, while $|\vec{\bm p}|$ and $d\Omega$ are the momentum magnitude
and solid-angle element in its rest frame. After performing the $\phi_{\rm CM}$ integration, one obtains
\begin{align}
d\Phi_4={}&\frac{Y\,Y_{pp}\,Y_{ee}}{2^{13}\pi^7\sqrt{s}\,m_{pp}m_{ee}}\,dm_{pp}^2\,dm_{ee}^2\,d\cos\theta_{\rm CM}d\cos\theta_{pp}\,d\phi_{pp}\,d\cos\theta_{ee}\,d\phi_{ee}.
\label{eq:dPhi4}
\end{align}

There are seven independent variables, and their limits are
\begin{align}
4m_p^2 \le m_{pp}^2\le(\sqrt{s}-2m_e)^2,  \qquad\,      4m_e^2 \le m_{ee}^2\le\left(\sqrt{s}-\sqrt{m_{pp}^2}\right)^2 ,
\label{eq:masslimits}
\end{align}
\begin{equation}
-1\le\cos\theta_{\rm CM},\cos\theta_{pp},\cos\theta_{ee}\le1, \qquad   0\le\phi_{pp},\phi_{ee}<2\pi.
\label{eq:angularlimits}
\end{equation}
In addition, the physical scattering threshold is
\begin{equation}
\sqrt{s}\ge\max\left\{m_{B_1}+m_{B_2},\,2m_p+2m_e\right\}.
\label{eq:threshold}
\end{equation}

The differential cross section is
\begin{equation}
d\sigma = \frac{1}{2!\,2!}\,\frac{\overline{|{\cal M}|^2}}{2\sqrt{\lambda(s,m_{B_1}^2,m_{B_2}^2)}}\,d\Phi_4 .
\label{eq:dsigmaMaster}
\end{equation}
The factor $1/(2!2!)$ removes the double counting associated with the two identical protons and the two identical electrons.

\section{\hspace{-0.6cm} Complete numerical cross sections} \label{AppendixB}

The full cross-section tables used in Sec.~\ref{sec:numerical} are collected here.
Tabs.~\ref{tab:NN_Dq2_combined}--\ref{tab:SS_Dq2_combined} contain the light-neutrino results and cutoff variations.
Tabs.~\ref{tab:HN_NN_NL}--\ref{tab:HN_SL_SS} contain the heavy-neutrino results at the two parameter values.
All cross sections are in $\mathrm{GeV}^{-2}$.
{\small
\renewcommand{\baselinestretch}{1.35}
\begin{table*}[b]
\centering
\caption{Cross sections  (in $\mathrm{GeV}^{-2}$) for $nn\to pp e^-e^-$ for five values of $m_{\beta\beta}$: $1$, $10$, $30$, $50$, and $100~\mathrm{meV}$. The values at $\Delta q^2=1.0\times10^{-2}\,\mathrm{GeV}^2$ are shown with upper and lower variations obtained with $\Delta q^2=3.0\times10^{-3}$ and $3.0\times10^{-2}\,\mathrm{GeV}^2$, respectively. These variations show only the cutoff dependence and are not statistical uncertainties. The same convention is used in Tabs.~\ref{tab:LL_Dq2_combined}--\ref{tab:SS_Dq2_combined}.}
\label{tab:NN_Dq2_combined}
\renewcommand{\arraystretch}{1.25}
\setlength{\tabcolsep}{8pt}
\resizebox{\textwidth}{!}{%
\begin{tabular}{cccccc}
\hline\hline
\rule[-1.2ex]{0pt}{5.0ex}
\shortstack{$\sqrt{s}$\\{(GeV)}}
& \shortstack{$1.0\times10^{-12}\,\mathrm{GeV}$\\$(1\,\mathrm{meV})$}  & \shortstack{$1.0\times10^{-11}\,\mathrm{GeV}$\\$(10\,\mathrm{meV})$}  & \shortstack{$3.0\times10^{-11}\,\mathrm{GeV}$\\$(30\,\mathrm{meV})$}  & \shortstack{$5.0\times10^{-11}\,\mathrm{GeV}$\\$(50\,\mathrm{meV})$}  & \shortstack{$1.0\times10^{-10}\,\mathrm{GeV}$\\$(100\,\mathrm{meV})$} \\\hline
1.9    &  $\left(2.38^{+1.65}_{-2.25}\right)\times10^{-55}$  &  $\left(2.38^{+1.65}_{-2.25}\right)\times10^{-53}$  &  $\left(2.14^{+1.48}_{-2.03}\right)\times10^{-52}$  &  $\left(5.95^{+4.12}_{-5.63}\right)\times10^{-52}$  &  $\left(2.38^{+1.65}_{-2.25}\right)\times10^{-51}$ \\ \hline
2.1    &  $\left(1.33^{+0.04}_{-0.21}\right)\times10^{-51}$  &  $\left(1.33^{+0.04}_{-0.21}\right)\times10^{-49}$  &  $\left(1.20^{+0.04}_{-0.19}\right)\times10^{-48}$  &  $\left(3.33^{+0.10}_{-0.52}\right)\times10^{-48}$  &  $\left(1.33^{+0.04}_{-0.21}\right)\times10^{-47}$ \\ \hline
2.3    &  $\left(1.06^{+0.02}_{-0.12}\right)\times10^{-50}$  &  $\left(1.06^{+0.02}_{-0.12}\right)\times10^{-48}$  &  $\left(9.58^{+0.19}_{-1.09}\right)\times10^{-48}$  &  $\left(2.66^{+0.05}_{-0.30}\right)\times10^{-47}$  &  $\left(1.06^{+0.02}_{-0.12}\right)\times10^{-46}$ \\ \hline
2.5    &  $\left(3.45^{+0.06}_{-0.35}\right)\times10^{-50}$  &  $\left(3.45^{+0.06}_{-0.35}\right)\times10^{-48}$  &  $\left(3.11^{+0.06}_{-0.31}\right)\times10^{-47}$  &  $\left(8.63^{+0.16}_{-0.87}\right)\times10^{-47}$  &  $\left(3.45^{+0.06}_{-0.35}\right)\times10^{-46}$ \\ \hline
2.7    &  $\left(7.66^{+0.17}_{-0.71}\right)\times10^{-50}$  &  $\left(7.66^{+0.17}_{-0.71}\right)\times10^{-48}$  &  $\left(6.89^{+0.16}_{-0.64}\right)\times10^{-47}$  &  $\left(1.91^{+0.04}_{-0.18}\right)\times10^{-46}$  &  $\left(7.66^{+0.17}_{-0.71}\right)\times10^{-46}$ \\ \hline
2.9    &  $\left(1.40^{+0.03}_{-0.13}\right)\times10^{-49}$  &  $\left(1.40^{+0.03}_{-0.13}\right)\times10^{-47}$  &  $\left(1.26^{+0.03}_{-0.12}\right)\times10^{-46}$  &  $\left(3.49^{+0.08}_{-0.32}\right)\times10^{-46}$  &  $\left(1.40^{+0.03}_{-0.13}\right)\times10^{-45}$ \\ \hline
3.1    &  $\left(2.23^{+0.07}_{-0.22}\right)\times10^{-49}$  &  $\left(2.23^{+0.07}_{-0.22}\right)\times10^{-47}$  &  $\left(2.01^{+0.06}_{-0.20}\right)\times10^{-46}$  &  $\left(5.59^{+0.17}_{-0.55}\right)\times10^{-46}$  &  $\left(2.23^{+0.07}_{-0.22}\right)\times10^{-45}$ \\ \hline
3.3    &  $\left(3.28^{+0.08}_{-0.31}\right)\times10^{-49}$  &  $\left(3.28^{+0.08}_{-0.31}\right)\times10^{-47}$  &  $\left(2.95^{+0.07}_{-0.28}\right)\times10^{-46}$  &  $\left(8.20^{+0.20}_{-0.78}\right)\times10^{-46}$  &  $\left(3.28^{+0.08}_{-0.31}\right)\times10^{-45}$ \\ \hline
3.5    &  $\left(4.52^{+0.13}_{-0.43}\right)\times10^{-49}$  &  $\left(4.52^{+0.13}_{-0.43}\right)\times10^{-47}$  &  $\left(4.07^{+0.12}_{-0.39}\right)\times10^{-46}$  &  $\left(1.13^{+0.03}_{-0.11}\right)\times10^{-45}$  &  $\left(4.52^{+0.13}_{-0.43}\right)\times10^{-45}$ \\ \hline
3.7    &  $\left(5.93^{+0.13}_{-0.58}\right)\times10^{-49}$  &  $\left(5.93^{+0.13}_{-0.58}\right)\times10^{-47}$  &  $\left(5.34^{+0.11}_{-0.52}\right)\times10^{-46}$  &  $\left(1.48^{+0.03}_{-0.15}\right)\times10^{-45}$  &  $\left(5.93^{+0.13}_{-0.58}\right)\times10^{-45}$ \\ \hline
3.9    &  $\left(7.51^{+0.30}_{-0.74}\right)\times10^{-49}$  &  $\left(7.51^{+0.30}_{-0.74}\right)\times10^{-47}$  &  $\left(6.76^{+0.27}_{-0.67}\right)\times10^{-46}$  &  $\left(1.88^{+0.08}_{-0.19}\right)\times10^{-45}$  &  $\left(7.51^{+0.30}_{-0.74}\right)\times10^{-45}$ \\ \hline
4.1    &  $\left(9.16^{+0.38}_{-0.88}\right)\times10^{-49}$  &  $\left(9.16^{+0.38}_{-0.88}\right)\times10^{-47}$  &  $\left(8.25^{+0.34}_{-0.79}\right)\times10^{-46}$  &  $\left(2.29^{+0.10}_{-0.22}\right)\times10^{-45}$  &  $\left(9.16^{+0.38}_{-0.88}\right)\times10^{-45}$ \\ \hline
4.3    &  $\left(1.11^{+0.04}_{-0.11}\right)\times10^{-48}$  &  $\left(1.11^{+0.04}_{-0.11}\right)\times10^{-46}$  &  $\left(9.95^{+0.38}_{-1.00}\right)\times10^{-46}$  &  $\left(2.76^{+0.11}_{-0.28}\right)\times10^{-45}$  &  $\left(1.11^{+0.04}_{-0.11}\right)\times10^{-44}$ \\ \hline
4.5    &  $\left(1.31^{+0.04}_{-0.14}\right)\times10^{-48}$  &  $\left(1.31^{+0.04}_{-0.14}\right)\times10^{-46}$  &  $\left(1.18^{+0.04}_{-0.12}\right)\times10^{-45}$  &  $\left(3.27^{+0.10}_{-0.35}\right)\times10^{-45}$  &  $\left(1.31^{+0.04}_{-0.14}\right)\times10^{-44}$ \\ \hline
4.7    &  $\left(1.51^{+0.05}_{-0.18}\right)\times10^{-48}$  &  $\left(1.51^{+0.05}_{-0.18}\right)\times10^{-46}$  &  $\left(1.36^{+0.05}_{-0.16}\right)\times10^{-45}$  &  $\left(3.78^{+0.13}_{-0.45}\right)\times10^{-45}$  &  $\left(1.51^{+0.05}_{-0.18}\right)\times10^{-44}$ \\ \hline
4.9    &  $\left(1.73^{+0.07}_{-0.18}\right)\times10^{-48}$  &  $\left(1.73^{+0.07}_{-0.18}\right)\times10^{-46}$  &  $\left(1.56^{+0.07}_{-0.16}\right)\times10^{-45}$  &  $\left(4.32^{+0.18}_{-0.44}\right)\times10^{-45}$  &  $\left(1.73^{+0.07}_{-0.18}\right)\times10^{-44}$ \\ \hline
5.1    &  $\left(1.98^{+0.06}_{-0.23}\right)\times10^{-48}$  &  $\left(1.98^{+0.06}_{-0.23}\right)\times10^{-46}$  &  $\left(1.78^{+0.06}_{-0.21}\right)\times10^{-45}$  &  $\left(4.94^{+0.15}_{-0.58}\right)\times10^{-45}$  &  $\left(1.98^{+0.06}_{-0.23}\right)\times10^{-44}$ \\ \hline
\end{tabular}}
%\end{table*}
%
%\begin{table*}[htbp]
\vspace{0.5cm}
 \centering
 \caption{Cross sections  (in $\mathrm{GeV}^{-2}$) for $\Lambda\Lambda\to pp e^-e^-$ for different values of $m_{\beta\beta}$. }
 \label{tab:LL_Dq2_combined}
 \renewcommand{\arraystretch}{1.25}
 \setlength{\tabcolsep}{8pt}
 \resizebox{\textwidth}{!}{%
 \begin{tabular}{cccccc}   \hline\hline   \rule[-1.2ex]{0pt}{5.0ex}
 \shortstack{$\sqrt{s}$\\{(GeV)}}    & \shortstack{$1.0\times10^{-12}\,\mathrm{GeV}$\\$(1\,\mathrm{meV})$}     & \shortstack{$1.0\times10^{-11}\,\mathrm{GeV}$\\$(10\,\mathrm{meV})$}      & \shortstack{$3.0\times10^{-11}\,\mathrm{GeV}$\\$(30\,\mathrm{meV})$}
 & \shortstack{$5.0\times10^{-11}\,\mathrm{GeV}$\\$(50\,\mathrm{meV})$}       & \shortstack{$1.0\times10^{-10}\,\mathrm{GeV}$\\$(100\,\mathrm{meV})$} \\      \hline
  2.3    &    $\left(3.00^{+5.88}_{-1.96}\right)\times10^{-51}$    &    $\left(3.00^{+5.88}_{-1.96}\right)\times10^{-49}$    &    $\left(2.70^{+5.29}_{-1.76}\right)\times10^{-48}$    &    $\left(7.49^{+14.70}_{-4.89}\right)\times10^{-48}$    &    $\left(3.00^{+5.88}_{-1.96}\right)\times10^{-47}$ \\\hline
  2.5    &    $\left(2.87^{+4.28}_{-1.67}\right)\times10^{-51}$    &    $\left(2.87^{+4.28}_{-1.67}\right)\times10^{-49}$    &    $\left(2.59^{+3.85}_{-1.50}\right)\times10^{-48}$    &    $\left(7.18^{+10.70}_{-4.18}\right)\times10^{-48}$    &    $\left(2.87^{+4.28}_{-1.67}\right)\times10^{-47}$ \\\hline
  2.7    &    $\left(3.64^{+4.22}_{-1.88}\right)\times10^{-51}$    &    $\left(3.64^{+4.22}_{-1.88}\right)\times10^{-49}$    &    $\left(3.28^{+3.80}_{-1.70}\right)\times10^{-48}$    &    $\left(9.10^{+10.55}_{-4.71}\right)\times10^{-48}$    &    $\left(3.64^{+4.22}_{-1.88}\right)\times10^{-47}$ \\\hline
  2.9    &    $\left(4.58^{+4.69}_{-2.12}\right)\times10^{-51}$    &    $\left(4.58^{+4.69}_{-2.12}\right)\times10^{-49}$    &    $\left(4.12^{+4.22}_{-1.91}\right)\times10^{-48}$    &    $\left(1.15^{+1.17}_{-0.53}\right)\times10^{-47}$    &    $\left(4.58^{+4.69}_{-2.12}\right)\times10^{-47}$ \\\hline
  3.1    &    $\left(5.83^{+5.01}_{-2.59}\right)\times10^{-51}$    &    $\left(5.83^{+5.01}_{-2.59}\right)\times10^{-49}$    &    $\left(5.25^{+4.50}_{-2.33}\right)\times10^{-48}$    &    $\left(1.46^{+1.25}_{-0.65}\right)\times10^{-47}$    &    $\left(5.83^{+5.01}_{-2.59}\right)\times10^{-47}$ \\\hline
  3.3    &    $\left(6.97^{+5.50}_{-2.83}\right)\times10^{-51}$    &    $\left(6.97^{+5.50}_{-2.83}\right)\times10^{-49}$    &    $\left(6.27^{+4.95}_{-2.55}\right)\times10^{-48}$    &    $\left(1.74^{+1.37}_{-0.71}\right)\times10^{-47}$    &    $\left(6.97^{+5.50}_{-2.83}\right)\times10^{-47}$ \\\hline
  3.5    &    $\left(8.20^{+5.90}_{-3.17}\right)\times10^{-51}$    &    $\left(8.20^{+5.90}_{-3.17}\right)\times10^{-49}$    &    $\left(7.38^{+5.31}_{-2.85}\right)\times10^{-48}$    &    $\left(2.05^{+1.47}_{-0.79}\right)\times10^{-47}$    &    $\left(8.20^{+5.90}_{-3.17}\right)\times10^{-47}$ \\\hline
  3.7    &    $\left(9.48^{+6.21}_{-3.39}\right)\times10^{-51}$    &    $\left(9.48^{+6.21}_{-3.39}\right)\times10^{-49}$    &    $\left(8.53^{+5.59}_{-3.06}\right)\times10^{-48}$    &    $\left(2.37^{+1.55}_{-0.85}\right)\times10^{-47}$    &    $\left(9.48^{+6.21}_{-3.39}\right)\times10^{-47}$ \\\hline
  3.9    &    $\left(1.08^{+0.67}_{-0.37}\right)\times10^{-50}$    &    $\left(1.08^{+0.67}_{-0.37}\right)\times10^{-48}$    &    $\left(9.75^{+6.07}_{-3.36}\right)\times10^{-48}$    &    $\left(2.71^{+1.69}_{-0.93}\right)\times10^{-47}$    &    $\left(1.08^{+0.67}_{-0.37}\right)\times10^{-46}$ \\\hline
  4.1    &    $\left(1.23^{+0.70}_{-0.42}\right)\times10^{-50}$    &    $\left(1.23^{+0.70}_{-0.42}\right)\times10^{-48}$    &    $\left(1.10^{+0.63}_{-0.38}\right)\times10^{-47}$    &    $\left(3.07^{+1.75}_{-1.04}\right)\times10^{-47}$    &    $\left(1.23^{+0.70}_{-0.42}\right)\times10^{-46}$ \\\hline
  4.3    &    $\left(1.37^{+0.72}_{-0.45}\right)\times10^{-50}$    &    $\left(1.37^{+0.72}_{-0.45}\right)\times10^{-48}$    &    $\left(1.23^{+0.65}_{-0.41}\right)\times10^{-47}$    &    $\left(3.42^{+1.81}_{-1.13}\right)\times10^{-47}$    &    $\left(1.37^{+0.72}_{-0.45}\right)\times10^{-46}$ \\\hline
  4.5    &    $\left(1.52^{+0.73}_{-0.49}\right)\times10^{-50}$    &    $\left(1.52^{+0.73}_{-0.49}\right)\times10^{-48}$    &    $\left(1.36^{+0.65}_{-0.44}\right)\times10^{-47}$    &    $\left(3.79^{+1.81}_{-1.22}\right)\times10^{-47}$    &    $\left(1.52^{+0.73}_{-0.49}\right)\times10^{-46}$ \\\hline
  4.7    &    $\left(1.65^{+0.80}_{-0.51}\right)\times10^{-50}$    &    $\left(1.65^{+0.80}_{-0.51}\right)\times10^{-48}$    &    $\left(1.48^{+0.72}_{-0.46}\right)\times10^{-47}$    &    $\left(4.12^{+2.00}_{-1.28}\right)\times10^{-47}$    &    $\left(1.65^{+0.80}_{-0.51}\right)\times10^{-46}$ \\\hline
  4.9    &    $\left(1.77^{+0.84}_{-0.53}\right)\times10^{-50}$    &    $\left(1.77^{+0.84}_{-0.53}\right)\times10^{-48}$    &    $\left(1.60^{+0.76}_{-0.47}\right)\times10^{-47}$    &    $\left(4.43^{+2.10}_{-1.32}\right)\times10^{-47}$    &    $\left(1.77^{+0.84}_{-0.53}\right)\times10^{-46}$ \\\hline
  5.1    &    $\left(1.91^{+0.91}_{-0.55}\right)\times10^{-50}$    &    $\left(1.91^{+0.91}_{-0.55}\right)\times10^{-48}$    &    $\left(1.72^{+0.82}_{-0.50}\right)\times10^{-47}$    &    $\left(4.77^{+2.28}_{-1.38}\right)\times10^{-47}$    &    $\left(1.91^{+0.91}_{-0.55}\right)\times10^{-46}$ \\\hline
 \end{tabular} }
\end{table*}	

\begin{table*}[htbp]
\centering
\caption{Cross sections  (in $\mathrm{GeV}^{-2}$) for $n\Lambda\to pp e^-e^-$ for different values of $m_{\beta\beta}$. }
\label{tab:NL_Dq2_combined}
\renewcommand{\arraystretch}{1.2}
\setlength{\tabcolsep}{8pt}
\resizebox{\textwidth}{!}{%
\begin{tabular}{cccccc}
\hline\hline
\rule[-1.2ex]{0pt}{5.0ex}
\shortstack{$\sqrt{s}$\\{(GeV)}}
& \shortstack{$1.0\times10^{-12}\,\mathrm{GeV}$\\$(1\,\mathrm{meV})$}  & \shortstack{$1.0\times10^{-11}\,\mathrm{GeV}$\\$(10\,\mathrm{meV})$}  & \shortstack{$3.0\times10^{-11}\,\mathrm{GeV}$\\$(30\,\mathrm{meV})$}  & \shortstack{$5.0\times10^{-11}\,\mathrm{GeV}$\\$(50\,\mathrm{meV})$}  & \shortstack{$1.0\times10^{-10}\,\mathrm{GeV}$\\$(100\,\mathrm{meV})$} \\\hline
2.1    &  $\left(3.31^{+7.25}_{-2.37}\right)\times10^{-51}$  &  $\left(3.31^{+7.25}_{-2.37}\right)\times10^{-49}$  &  $\left(2.98^{+6.53}_{-2.13}\right)\times10^{-48}$  &  $\left(8.27^{+18.13}_{-5.92}\right)\times10^{-48}$  &  $\left(3.31^{+7.25}_{-2.37}\right)\times10^{-47}$ \\ \hline
2.3    &  $\left(6.27^{+8.28}_{-3.34}\right)\times10^{-51}$  &  $\left(6.27^{+8.28}_{-3.34}\right)\times10^{-49}$  &  $\left(5.64^{+7.45}_{-3.00}\right)\times10^{-48}$  &  $\left(1.57^{+2.07}_{-0.83}\right)\times10^{-47}$   &  $\left(6.27^{+8.28}_{-3.34}\right)\times10^{-47}$ \\ \hline
2.5    &  $\left(1.21^{+1.20}_{-0.54}\right)\times10^{-50}$  &  $\left(1.21^{+1.20}_{-0.54}\right)\times10^{-48}$  &  $\left(1.09^{+1.08}_{-0.49}\right)\times10^{-47}$  &  $\left(3.02^{+3.01}_{-1.36}\right)\times10^{-47}$   &  $\left(1.21^{+1.20}_{-0.54}\right)\times10^{-46}$ \\ \hline
2.7    &  $\left(2.03^{+1.51}_{-0.82}\right)\times10^{-50}$  &  $\left(2.03^{+1.51}_{-0.82}\right)\times10^{-48}$  &  $\left(1.83^{+1.36}_{-0.74}\right)\times10^{-47}$  &  $\left(5.08^{+3.77}_{-2.05}\right)\times10^{-47}$   &  $\left(2.03^{+1.51}_{-0.82}\right)\times10^{-46}$ \\ \hline
2.9    &  $\left(2.98^{+1.95}_{-1.06}\right)\times10^{-50}$  &  $\left(2.98^{+1.95}_{-1.06}\right)\times10^{-48}$  &  $\left(2.68^{+1.75}_{-0.95}\right)\times10^{-47}$  &  $\left(7.45^{+4.87}_{-2.64}\right)\times10^{-47}$   &  $\left(2.98^{+1.95}_{-1.06}\right)\times10^{-46}$ \\ \hline
3.1    &  $\left(4.18^{+2.30}_{-1.43}\right)\times10^{-50}$  &  $\left(4.18^{+2.30}_{-1.43}\right)\times10^{-48}$  &  $\left(3.76^{+2.07}_{-1.28}\right)\times10^{-47}$  &  $\left(1.04^{+0.58}_{-0.36}\right)\times10^{-46}$   &  $\left(4.18^{+2.30}_{-1.43}\right)\times10^{-46}$ \\ \hline
3.3    &  $\left(5.47^{+2.74}_{-1.71}\right)\times10^{-50}$  &  $\left(5.47^{+2.74}_{-1.71}\right)\times10^{-48}$  &  $\left(4.92^{+2.46}_{-1.54}\right)\times10^{-47}$  &  $\left(1.37^{+0.68}_{-0.43}\right)\times10^{-46}$   &  $\left(5.47^{+2.74}_{-1.71}\right)\times10^{-46}$ \\ \hline
3.5    &  $\left(6.85^{+3.17}_{-1.99}\right)\times10^{-50}$  &  $\left(6.85^{+3.17}_{-1.99}\right)\times10^{-48}$  &  $\left(6.17^{+2.86}_{-1.79}\right)\times10^{-47}$  &  $\left(1.71^{+0.79}_{-0.50}\right)\times10^{-46}$   &  $\left(6.85^{+3.17}_{-1.99}\right)\times10^{-46}$ \\ \hline
3.7    &  $\left(8.43^{+3.29}_{-2.45}\right)\times10^{-50}$  &  $\left(8.43^{+3.29}_{-2.45}\right)\times10^{-48}$  &  $\left(7.59^{+2.96}_{-2.20}\right)\times10^{-47}$  &  $\left(2.11^{+0.82}_{-0.61}\right)\times10^{-46}$   &  $\left(8.43^{+3.29}_{-2.45}\right)\times10^{-46}$ \\ \hline
3.9    &  $\left(1.00^{+0.37}_{-0.27}\right)\times10^{-49}$  &  $\left(1.00^{+0.37}_{-0.27}\right)\times10^{-47}$  &  $\left(9.02^{+3.34}_{-2.47}\right)\times10^{-47}$  &  $\left(2.51^{+0.93}_{-0.69}\right)\times10^{-46}$   &  $\left(1.00^{+0.37}_{-0.27}\right)\times10^{-45}$ \\ \hline
4.1    &  $\left(1.18^{+0.40}_{-0.32}\right)\times10^{-49}$  &  $\left(1.18^{+0.40}_{-0.32}\right)\times10^{-47}$  &  $\left(1.06^{+0.36}_{-0.29}\right)\times10^{-46}$  &  $\left(2.95^{+0.99}_{-0.80}\right)\times10^{-46}$   &  $\left(1.18^{+0.40}_{-0.32}\right)\times10^{-45}$ \\ \hline
4.3    &  $\left(1.34^{+0.46}_{-0.34}\right)\times10^{-49}$  &  $\left(1.34^{+0.46}_{-0.34}\right)\times10^{-47}$  &  $\left(1.21^{+0.41}_{-0.31}\right)\times10^{-46}$  &  $\left(3.35^{+1.15}_{-0.86}\right)\times10^{-46}$   &  $\left(1.34^{+0.46}_{-0.34}\right)\times10^{-45}$ \\ \hline
4.5    &  $\left(1.50^{+0.51}_{-0.37}\right)\times10^{-49}$  &  $\left(1.50^{+0.51}_{-0.37}\right)\times10^{-47}$  &  $\left(1.35^{+0.46}_{-0.33}\right)\times10^{-46}$  &  $\left(3.76^{+1.27}_{-0.92}\right)\times10^{-46}$   &  $\left(1.50^{+0.51}_{-0.37}\right)\times10^{-45}$ \\ \hline
4.7    &  $\left(1.68^{+0.58}_{-0.39}\right)\times10^{-49}$  &  $\left(1.68^{+0.58}_{-0.39}\right)\times10^{-47}$  &  $\left(1.51^{+0.53}_{-0.35}\right)\times10^{-46}$  &  $\left(4.20^{+1.46}_{-0.99}\right)\times10^{-46}$   &  $\left(1.68^{+0.58}_{-0.39}\right)\times10^{-45}$ \\ \hline
4.9    &  $\left(1.87^{+0.59}_{-0.43}\right)\times10^{-49}$  &  $\left(1.87^{+0.59}_{-0.43}\right)\times10^{-47}$  &  $\left(1.69^{+0.53}_{-0.38}\right)\times10^{-46}$  &  $\left(4.68^{+1.47}_{-1.07}\right)\times10^{-46}$   &  $\left(1.87^{+0.59}_{-0.43}\right)\times10^{-45}$ \\ \hline
5.1    &  $\left(2.05^{+0.63}_{-0.47}\right)\times10^{-49}$  &  $\left(2.05^{+0.63}_{-0.47}\right)\times10^{-47}$  &  $\left(1.84^{+0.56}_{-0.42}\right)\times10^{-46}$  &  $\left(5.12^{+1.57}_{-1.18}\right)\times10^{-46}$   &  $\left(2.05^{+0.63}_{-0.47}\right)\times10^{-45}$ \\ \hline
\end{tabular}}
%\end{table*}
%
%\begin{table*}[htbp]
\vspace{1cm}
\centering
\caption{Cross sections  (in $\mathrm{GeV}^{-2}$) for $n\Sigma^0\to pp e^-e^-$ for different values of	$m_{\beta\beta}$. }
\label{tab:NS_Dq2_combined}
\renewcommand{\arraystretch}{1.25}
\setlength{\tabcolsep}{8pt}
\resizebox{\textwidth}{!}{%
\begin{tabular}{cccccc}
\hline\hline
\rule[-1.2ex]{0pt}{5.0ex}
\shortstack{$\sqrt{s}$\\{(GeV)}}& \shortstack{$1.0\times10^{-12}\,\mathrm{GeV}$\\$(1\,\mathrm{meV})$}  & \shortstack{$1.0\times10^{-11}\,\mathrm{GeV}$\\$(10\,\mathrm{meV})$}  & \shortstack{$3.0\times10^{-11}\,\mathrm{GeV}$\\$(30\,\mathrm{meV})$}  & \shortstack{$5.0\times10^{-11}\,\mathrm{GeV}$\\$(50\,\mathrm{meV})$}  & \shortstack{$1.0\times10^{-10}\,\mathrm{GeV}$\\$(100\,\mathrm{meV})$} \\\hline
2.15   &  $\left(2.83^{+10.09}_{-2.30}\right)\times10^{-51}$  &  $\left(2.83^{+10.09}_{-2.30}\right)\times10^{-49}$  &  $\left(2.54^{+9.08}_{-2.07}\right)\times10^{-48}$   &  $\left(7.06^{+25.23}_{-5.74}\right)\times10^{-48}$  &  $\left(2.83^{+10.09}_{-2.30}\right)\times10^{-47}$ \\ \hline
2.3    &  $\left(2.94^{+6.85}_{-2.04}\right)\times10^{-51}$   &  $\left(2.94^{+6.85}_{-2.04}\right)\times10^{-49}$   &  $\left(2.64^{+6.17}_{-1.83}\right)\times10^{-48}$   &  $\left(7.34^{+17.13}_{-5.09}\right)\times10^{-48}$  &  $\left(2.94^{+6.85}_{-2.04}\right)\times10^{-47}$  \\ \hline
2.5    &  $\left(4.87^{+8.50}_{-2.92}\right)\times10^{-51}$   &  $\left(4.87^{+8.50}_{-2.92}\right)\times10^{-49}$   &  $\left(4.38^{+7.65}_{-2.63}\right)\times10^{-48}$   &  $\left(1.22^{+2.13}_{-0.73}\right)\times10^{-47}$   &  $\left(4.87^{+8.50}_{-2.92}\right)\times10^{-47}$  \\ \hline
2.7    &  $\left(7.32^{+10.86}_{-3.92}\right)\times10^{-51}$  &  $\left(7.32^{+10.86}_{-3.92}\right)\times10^{-49}$  &  $\left(6.59^{+9.77}_{-3.53}\right)\times10^{-48}$   &  $\left(1.83^{+2.71}_{-0.98}\right)\times10^{-47}$   &  $\left(7.32^{+10.86}_{-3.92}\right)\times10^{-47}$ \\ \hline
2.9    &  $\left(1.02^{+1.30}_{-0.50}\right)\times10^{-50}$   &  $\left(1.02^{+1.30}_{-0.50}\right)\times10^{-48}$   &  $\left(9.14^{+11.74}_{-4.50}\right)\times10^{-48}$  &  $\left(2.54^{+3.26}_{-1.25}\right)\times10^{-47}$   &  $\left(1.02^{+1.30}_{-0.50}\right)\times10^{-46}$  \\ \hline
3.1    &  $\left(1.33^{+1.49}_{-0.61}\right)\times10^{-50}$   &  $\left(1.33^{+1.49}_{-0.61}\right)\times10^{-48}$   &  $\left(1.19^{+1.34}_{-0.54}\right)\times10^{-47}$   &  $\left(3.32^{+3.73}_{-1.51}\right)\times10^{-47}$   &  $\left(1.33^{+1.49}_{-0.61}\right)\times10^{-46}$  \\ \hline
3.3    &  $\left(1.66^{+1.67}_{-0.71}\right)\times10^{-50}$   &  $\left(1.66^{+1.67}_{-0.71}\right)\times10^{-48}$   &  $\left(1.50^{+1.51}_{-0.64}\right)\times10^{-47}$   &  $\left(4.15^{+4.18}_{-1.77}\right)\times10^{-47}$   &  $\left(1.66^{+1.67}_{-0.71}\right)\times10^{-46}$  \\ \hline
3.5    &  $\left(2.00^{+1.82}_{-0.81}\right)\times10^{-50}$   &  $\left(2.00^{+1.82}_{-0.81}\right)\times10^{-48}$   &  $\left(1.80^{+1.64}_{-0.73}\right)\times10^{-47}$   &  $\left(5.01^{+4.54}_{-2.03}\right)\times10^{-47}$   &  $\left(2.00^{+1.82}_{-0.81}\right)\times10^{-46}$  \\ \hline
3.7    &  $\left(2.36^{+2.03}_{-0.89}\right)\times10^{-50}$   &  $\left(2.36^{+2.03}_{-0.89}\right)\times10^{-48}$   &  $\left(2.12^{+1.83}_{-0.80}\right)\times10^{-47}$   &  $\left(5.89^{+5.08}_{-2.23}\right)\times10^{-47}$   &  $\left(2.36^{+2.03}_{-0.89}\right)\times10^{-46}$  \\ \hline
3.9    &  $\left(2.72^{+2.11}_{-0.98}\right)\times10^{-50}$   &  $\left(2.72^{+2.11}_{-0.98}\right)\times10^{-48}$   &  $\left(2.45^{+1.90}_{-0.88}\right)\times10^{-47}$   &  $\left(6.80^{+5.27}_{-2.44}\right)\times10^{-47}$   &  $\left(2.72^{+2.11}_{-0.98}\right)\times10^{-46}$  \\ \hline
4.1    &  $\left(3.10^{+2.41}_{-1.06}\right)\times10^{-50}$   &  $\left(3.10^{+2.41}_{-1.06}\right)\times10^{-48}$   &  $\left(2.79^{+2.17}_{-0.96}\right)\times10^{-47}$   &  $\left(7.74^{+6.03}_{-2.66}\right)\times10^{-47}$   &  $\left(3.10^{+2.41}_{-1.06}\right)\times10^{-46}$  \\ \hline
4.3    &  $\left(3.53^{+2.38}_{-1.18}\right)\times10^{-50}$   &  $\left(3.53^{+2.38}_{-1.18}\right)\times10^{-48}$   &  $\left(3.18^{+2.14}_{-1.06}\right)\times10^{-47}$   &  $\left(8.83^{+5.94}_{-2.96}\right)\times10^{-47}$   &  $\left(3.53^{+2.38}_{-1.18}\right)\times10^{-46}$  \\ \hline
4.5    &  $\left(3.98^{+2.33}_{-1.36}\right)\times10^{-50}$   &  $\left(3.98^{+2.33}_{-1.36}\right)\times10^{-48}$   &  $\left(3.58^{+2.10}_{-1.22}\right)\times10^{-47}$   &  $\left(9.95^{+5.82}_{-3.39}\right)\times10^{-47}$   &  $\left(3.98^{+2.33}_{-1.36}\right)\times10^{-46}$  \\ \hline
4.7    &  $\left(4.27^{+2.65}_{-1.31}\right)\times10^{-50}$   &  $\left(4.27^{+2.65}_{-1.31}\right)\times10^{-48}$   &  $\left(3.85^{+2.38}_{-1.18}\right)\times10^{-47}$   &  $\left(1.07^{+0.66}_{-0.33}\right)\times10^{-46}$   &  $\left(4.27^{+2.65}_{-1.31}\right)\times10^{-46}$  \\ \hline
4.9    &  $\left(4.70^{+2.88}_{-1.42}\right)\times10^{-50}$   &  $\left(4.70^{+2.88}_{-1.42}\right)\times10^{-48}$   &  $\left(4.23^{+2.60}_{-1.28}\right)\times10^{-47}$   &  $\left(1.18^{+0.72}_{-0.35}\right)\times10^{-46}$   &  $\left(4.70^{+2.88}_{-1.42}\right)\times10^{-46}$  \\ \hline
5.1    &  $\left(5.16^{+2.81}_{-1.53}\right)\times10^{-50}$   &  $\left(5.16^{+2.81}_{-1.53}\right)\times10^{-48}$   &  $\left(4.64^{+2.53}_{-1.38}\right)\times10^{-47}$   &  $\left(1.29^{+0.70}_{-0.38}\right)\times10^{-46}$   &  $\left(5.16^{+2.81}_{-1.53}\right)\times10^{-46}$  \\ \hline
\end{tabular}}
\end{table*}
\begin{table*}[htbp]
\centering
\caption{Cross sections  (in $\mathrm{GeV}^{-2}$) for $\Lambda\Sigma^0\to pp e^-e^-$ for different values of	$m_{\beta\beta}$. }
\label{tab:SL_Dq2_combined}
\renewcommand{\arraystretch}{1.3}
\setlength{\tabcolsep}{8pt}
\resizebox{\textwidth}{!}{%
\begin{tabular}{cccccc}
\hline\hline
\rule[-1.2ex]{0pt}{5.0ex}
\shortstack{$\sqrt{s}$\\{(GeV)}}
& \shortstack{$1.0\times10^{-12}\,\mathrm{GeV}$\\$(1\,\mathrm{meV})$}  & \shortstack{$1.0\times10^{-11}\,\mathrm{GeV}$\\$(10\,\mathrm{meV})$}  & \shortstack{$3.0\times10^{-11}\,\mathrm{GeV}$\\$(30\,\mathrm{meV})$}  & \shortstack{$5.0\times10^{-11}\,\mathrm{GeV}$\\$(50\,\mathrm{meV})$}  & \shortstack{$1.0\times10^{-10}\,\mathrm{GeV}$\\$(100\,\mathrm{meV})$} \\
\hline
2.32   &  $\left(1.84^{+5.34}_{-1.42}\right)\times10^{-51}$  &  $\left(1.84^{+5.34}_{-1.42}\right)\times10^{-49}$  &  $\left(1.66^{+4.81}_{-1.27}\right)\times10^{-48}$   &  $\left(4.61^{+13.35}_{-3.54}\right)\times10^{-48}$  &  $\left(1.84^{+5.34}_{-1.42}\right)\times10^{-47}$ \\ \hline
2.5    &  $\left(1.00^{+1.89}_{-0.66}\right)\times10^{-51}$  &  $\left(1.00^{+1.89}_{-0.66}\right)\times10^{-49}$  &  $\left(9.04^{+16.99}_{-5.97}\right)\times10^{-49}$  &  $\left(2.51^{+4.72}_{-1.66}\right)\times10^{-48}$   &  $\left(1.00^{+1.89}_{-0.66}\right)\times10^{-47}$ \\ \hline
2.7    &  $\left(1.13^{+1.80}_{-0.67}\right)\times10^{-51}$  &  $\left(1.13^{+1.80}_{-0.67}\right)\times10^{-49}$  &  $\left(1.02^{+1.62}_{-0.60}\right)\times10^{-48}$   &  $\left(2.83^{+4.49}_{-1.67}\right)\times10^{-48}$   &  $\left(1.13^{+1.80}_{-0.67}\right)\times10^{-47}$ \\ \hline
2.9    &  $\left(1.33^{+1.81}_{-0.71}\right)\times10^{-51}$  &  $\left(1.33^{+1.81}_{-0.71}\right)\times10^{-49}$  &  $\left(1.20^{+1.63}_{-0.64}\right)\times10^{-48}$   &  $\left(3.33^{+4.53}_{-1.78}\right)\times10^{-48}$   &  $\left(1.33^{+1.81}_{-0.71}\right)\times10^{-47}$ \\ \hline
3.1    &  $\left(1.58^{+1.94}_{-0.78}\right)\times10^{-51}$  &  $\left(1.58^{+1.94}_{-0.78}\right)\times10^{-49}$  &  $\left(1.42^{+1.74}_{-0.70}\right)\times10^{-48}$   &  $\left(3.94^{+4.84}_{-1.95}\right)\times10^{-48}$   &  $\left(1.58^{+1.94}_{-0.78}\right)\times10^{-47}$ \\ \hline
3.3    &  $\left(1.83^{+2.10}_{-0.85}\right)\times10^{-51}$  &  $\left(1.83^{+2.10}_{-0.85}\right)\times10^{-49}$  &  $\left(1.65^{+1.89}_{-0.76}\right)\times10^{-48}$   &  $\left(4.58^{+5.25}_{-2.12}\right)\times10^{-48}$   &  $\left(1.83^{+2.10}_{-0.85}\right)\times10^{-47}$ \\ \hline
3.5    &  $\left(2.12^{+2.14}_{-0.94}\right)\times10^{-51}$  &  $\left(2.12^{+2.14}_{-0.94}\right)\times10^{-49}$  &  $\left(1.91^{+1.92}_{-0.85}\right)\times10^{-48}$   &  $\left(5.30^{+5.35}_{-2.35}\right)\times10^{-48}$   &  $\left(2.12^{+2.14}_{-0.94}\right)\times10^{-47}$ \\ \hline
3.7    &  $\left(2.43^{+2.23}_{-1.05}\right)\times10^{-51}$  &  $\left(2.43^{+2.23}_{-1.05}\right)\times10^{-49}$  &  $\left(2.18^{+2.01}_{-0.94}\right)\times10^{-48}$   &  $\left(6.06^{+5.57}_{-2.61}\right)\times10^{-48}$   &  $\left(2.43^{+2.23}_{-1.05}\right)\times10^{-47}$ \\ \hline
3.9    &  $\left(2.68^{+2.31}_{-1.06}\right)\times10^{-51}$  &  $\left(2.68^{+2.31}_{-1.06}\right)\times10^{-49}$  &  $\left(2.41^{+2.08}_{-0.96}\right)\times10^{-48}$   &  $\left(6.70^{+5.76}_{-2.66}\right)\times10^{-48}$   &  $\left(2.68^{+2.31}_{-1.06}\right)\times10^{-47}$ \\ \hline
4.1    &  $\left(2.92^{+2.42}_{-1.08}\right)\times10^{-51}$  &  $\left(2.92^{+2.42}_{-1.08}\right)\times10^{-49}$  &  $\left(2.62^{+2.18}_{-0.97}\right)\times10^{-48}$   &  $\left(7.29^{+6.05}_{-2.69}\right)\times10^{-48}$   &  $\left(2.92^{+2.42}_{-1.08}\right)\times10^{-47}$ \\ \hline
4.3    &  $\left(3.29^{+2.38}_{-1.25}\right)\times10^{-51}$  &  $\left(3.29^{+2.38}_{-1.25}\right)\times10^{-49}$  &  $\left(2.96^{+2.15}_{-1.12}\right)\times10^{-48}$   &  $\left(8.22^{+5.96}_{-3.11}\right)\times10^{-48}$   &  $\left(3.29^{+2.38}_{-1.25}\right)\times10^{-47}$ \\ \hline
4.5    &  $\left(3.55^{+2.60}_{-1.28}\right)\times10^{-51}$  &  $\left(3.55^{+2.60}_{-1.28}\right)\times10^{-49}$  &  $\left(3.19^{+2.34}_{-1.15}\right)\times10^{-48}$   &  $\left(8.87^{+6.51}_{-3.19}\right)\times10^{-48}$   &  $\left(3.55^{+2.60}_{-1.28}\right)\times10^{-47}$ \\ \hline
4.7    &  $\left(3.81^{+2.62}_{-1.31}\right)\times10^{-51}$  &  $\left(3.81^{+2.62}_{-1.31}\right)\times10^{-49}$  &  $\left(3.43^{+2.36}_{-1.18}\right)\times10^{-48}$   &  $\left(9.53^{+6.56}_{-3.28}\right)\times10^{-48}$   &  $\left(3.81^{+2.62}_{-1.31}\right)\times10^{-47}$ \\ \hline
4.9    &  $\left(4.12^{+2.79}_{-1.39}\right)\times10^{-51}$  &  $\left(4.12^{+2.79}_{-1.39}\right)\times10^{-49}$  &  $\left(3.71^{+2.51}_{-1.25}\right)\times10^{-48}$   &  $\left(1.03^{+0.70}_{-0.35}\right)\times10^{-47}$   &  $\left(4.12^{+2.79}_{-1.39}\right)\times10^{-47}$ \\ \hline
5.1    &  $\left(4.41^{+2.92}_{-1.45}\right)\times10^{-51}$  &  $\left(4.41^{+2.92}_{-1.45}\right)\times10^{-49}$  &  $\left(3.97^{+2.63}_{-1.30}\right)\times10^{-48}$   &  $\left(1.10^{+0.73}_{-0.36}\right)\times10^{-47}$   &  $\left(4.41^{+2.92}_{-1.45}\right)\times10^{-47}$ \\ \hline
\end{tabular}}
%\end{table*}
%	
%\begin{table*}[htbp]
\vspace{1cm}
\centering
\caption{Cross sections  (in $\mathrm{GeV}^{-2}$) for $\Sigma^0\Sigma^0\to pp e^-e^-$ for different values of	$m_{\beta\beta}$. }
\label{tab:SS_Dq2_combined}
\renewcommand{\arraystretch}{1.3}
\setlength{\tabcolsep}{8pt}
\resizebox{\textwidth}{!}{%
\begin{tabular}{cccccc}
\hline\hline
\rule[-1.2ex]{0pt}{5.0ex}\shortstack{$\sqrt{s}$\\{(GeV)}}& \shortstack{$1.0\times10^{-12}\,\mathrm{GeV}$\\$(1\,\mathrm{meV})$}  & \shortstack{$1.0\times10^{-11}\,\mathrm{GeV}$\\$(10\,\mathrm{meV})$}  & \shortstack{$3.0\times10^{-11}\,\mathrm{GeV}$\\$(30\,\mathrm{meV})$}  & \shortstack{$5.0\times10^{-11}\,\mathrm{GeV}$\\$(50\,\mathrm{meV})$}  & \shortstack{$1.0\times10^{-10}\,\mathrm{GeV}$\\$(100\,\mathrm{meV})$} \\\hline
2.39   &  $\left(1.62^{+2.97}_{-1.08}\right)\times10^{-51}$  &  $\left(1.62^{+2.97}_{-1.08}\right)\times10^{-49}$  &  $\left(1.46^{+2.67}_{-0.98}\right)\times10^{-48}$  &  $\left(4.04^{+7.42}_{-2.71}\right)\times10^{-48}$   &  $\left(1.62^{+2.97}_{-1.08}\right)\times10^{-47}$ \\ \hline
2.5    &  $\left(4.59^{+7.79}_{-2.95}\right)\times10^{-52}$  &  $\left(4.59^{+7.79}_{-2.95}\right)\times10^{-50}$  &  $\left(4.13^{+7.01}_{-2.65}\right)\times10^{-49}$  &  $\left(1.15^{+1.95}_{-0.74}\right)\times10^{-48}$   &  $\left(4.59^{+7.79}_{-2.95}\right)\times10^{-48}$ \\ \hline
2.7    &  $\left(3.94^{+5.69}_{-2.27}\right)\times10^{-52}$  &  $\left(3.94^{+5.69}_{-2.27}\right)\times10^{-50}$  &  $\left(3.55^{+5.12}_{-2.05}\right)\times10^{-49}$  &  $\left(9.86^{+14.23}_{-5.69}\right)\times10^{-49}$  &  $\left(3.94^{+5.69}_{-2.27}\right)\times10^{-48}$ \\ \hline
2.9    &  $\left(4.14^{+5.43}_{-2.17}\right)\times10^{-52}$  &  $\left(4.14^{+5.43}_{-2.17}\right)\times10^{-50}$  &  $\left(3.72^{+4.89}_{-1.96}\right)\times10^{-49}$  &  $\left(1.03^{+1.36}_{-0.54}\right)\times10^{-48}$   &  $\left(4.14^{+5.43}_{-2.17}\right)\times10^{-48}$ \\ \hline
3.1    &  $\left(4.53^{+5.41}_{-2.22}\right)\times10^{-52}$  &  $\left(4.53^{+5.41}_{-2.22}\right)\times10^{-50}$  &  $\left(4.08^{+4.87}_{-2.00}\right)\times10^{-49}$  &  $\left(1.13^{+1.35}_{-0.55}\right)\times10^{-48}$   &  $\left(4.53^{+5.41}_{-2.22}\right)\times10^{-48}$ \\ \hline
3.3    &  $\left(5.04^{+5.78}_{-2.33}\right)\times10^{-52}$  &  $\left(5.04^{+5.78}_{-2.33}\right)\times10^{-50}$  &  $\left(4.54^{+5.20}_{-2.10}\right)\times10^{-49}$  &  $\left(1.26^{+1.44}_{-0.58}\right)\times10^{-48}$   &  $\left(5.04^{+5.78}_{-2.33}\right)\times10^{-48}$ \\ \hline
3.5    &  $\left(5.59^{+5.74}_{-2.44}\right)\times10^{-52}$  &  $\left(5.59^{+5.74}_{-2.44}\right)\times10^{-50}$  &  $\left(5.03^{+5.16}_{-2.19}\right)\times10^{-49}$  &  $\left(1.40^{+1.43}_{-0.61}\right)\times10^{-48}$   &  $\left(5.59^{+5.74}_{-2.44}\right)\times10^{-48}$ \\ \hline
3.7    &  $\left(6.14^{+6.05}_{-2.57}\right)\times10^{-52}$  &  $\left(6.14^{+6.05}_{-2.57}\right)\times10^{-50}$  &  $\left(5.53^{+5.45}_{-2.32}\right)\times10^{-49}$  &  $\left(1.54^{+1.51}_{-0.64}\right)\times10^{-48}$   &  $\left(6.14^{+6.05}_{-2.57}\right)\times10^{-48}$ \\ \hline
3.9    &  $\left(6.74^{+6.13}_{-2.70}\right)\times10^{-52}$  &  $\left(6.74^{+6.13}_{-2.70}\right)\times10^{-50}$  &  $\left(6.07^{+5.51}_{-2.43}\right)\times10^{-49}$  &  $\left(1.69^{+1.53}_{-0.67}\right)\times10^{-48}$   &  $\left(6.74^{+6.13}_{-2.70}\right)\times10^{-48}$ \\ \hline
4.1    &  $\left(7.16^{+6.41}_{-2.63}\right)\times10^{-52}$  &  $\left(7.16^{+6.41}_{-2.63}\right)\times10^{-50}$  &  $\left(6.45^{+5.77}_{-2.36}\right)\times10^{-49}$  &  $\left(1.79^{+1.60}_{-0.66}\right)\times10^{-48}$   &  $\left(7.16^{+6.41}_{-2.63}\right)\times10^{-48}$ \\ \hline
4.3    &  $\left(7.89^{+6.46}_{-2.88}\right)\times10^{-52}$  &  $\left(7.89^{+6.46}_{-2.88}\right)\times10^{-50}$  &  $\left(7.10^{+5.81}_{-2.60}\right)\times10^{-49}$  &  $\left(1.97^{+1.62}_{-0.72}\right)\times10^{-48}$   &  $\left(7.89^{+6.46}_{-2.88}\right)\times10^{-48}$ \\ \hline
4.5    &  $\left(8.56^{+6.65}_{-3.10}\right)\times10^{-52}$  &  $\left(8.56^{+6.65}_{-3.10}\right)\times10^{-50}$  &  $\left(7.71^{+5.99}_{-2.79}\right)\times10^{-49}$  &  $\left(2.14^{+1.66}_{-0.77}\right)\times10^{-48}$   &  $\left(8.56^{+6.65}_{-3.10}\right)\times10^{-48}$ \\ \hline
4.7    &  $\left(9.14^{+7.03}_{-3.20}\right)\times10^{-52}$  &  $\left(9.14^{+7.03}_{-3.20}\right)\times10^{-50}$  &  $\left(8.22^{+6.33}_{-2.88}\right)\times10^{-49}$  &  $\left(2.28^{+1.76}_{-0.80}\right)\times10^{-48}$   &  $\left(9.14^{+7.03}_{-3.20}\right)\times10^{-48}$ \\ \hline
4.9    &  $\left(9.79^{+6.76}_{-3.33}\right)\times10^{-52}$  &  $\left(9.79^{+6.76}_{-3.33}\right)\times10^{-50}$  &  $\left(8.81^{+6.08}_{-3.00}\right)\times10^{-49}$  &  $\left(2.45^{+1.69}_{-0.83}\right)\times10^{-48}$   &  $\left(9.79^{+6.76}_{-3.33}\right)\times10^{-48}$ \\ \hline
5.1    &  $\left(1.03^{+0.70}_{-0.33}\right)\times10^{-51}$  &  $\left(1.03^{+0.70}_{-0.33}\right)\times10^{-49}$  &  $\left(9.28^{+6.32}_{-2.96}\right)\times10^{-49}$  &  $\left(2.58^{+1.76}_{-0.82}\right)\times10^{-48}$   &  $\left(1.03^{+0.70}_{-0.33}\right)\times10^{-47}$ \\ \hline
\end{tabular}	}
\end{table*}

}

\begin{table}[b]
\centering
\caption{Total cross sections (in $\mathrm{GeV}^{-2}$) for
$nn\to pp e^-e^-$ and $n\Lambda\to pp e^-e^-$ mediated by heavy
Majorana neutrino exchange.  The two parameters are
$\eta_N^A=4.0\times10^{-9}\,\mathrm{GeV}^{-1}$ and
$\eta_N^B=8.0\times10^{-9}\,\mathrm{GeV}^{-1}$.  Since
$\sigma\propto\eta_N^2$, the results for $\eta_N^B$ are exactly four times
those for $\eta_N^A$.  In this and the following two tables, a dash denotes an energy below
the corresponding channel threshold, while an ellipsis denotes an
energy point that was not calculated.}
\label{tab:HN_NN_NL}
\renewcommand{\arraystretch}{1.1}
\setlength{\tabcolsep}{25pt}
\begin{tabular}{c|cc|cc}
\hline\hline
\rule[-1.2ex]{0pt}{5.0ex}
\raisebox{-2.4ex}{\shortstack{$\sqrt{s}$\\{(GeV)}}}
     &       \multicolumn{2}{c|}{$nn\to pp e^-e^-$}
     &       \multicolumn{2}{c}{$n\Lambda\to pp e^-e^-$} \\
\cline{2-5}
     &       $\eta_N^A$      &       $\eta_N^B$      &       $\eta_N^A$      &       $\eta_N^B$ \\
\hline
1.9      &       $1.83\times10^{-51}$      &       $7.31\times10^{-51}$      &       --      &       -- \\ \hline
2.1      &       $3.83\times10^{-46}$      &       $1.53\times10^{-45}$      &       $7.63\times10^{-47}$      &       $3.05\times10^{-46}$ \\ \hline
2.3      &       $7.28\times10^{-45}$      &       $2.91\times10^{-44}$      &       $8.33\times10^{-46}$      &       $3.33\times10^{-45}$ \\ \hline
2.5      &       $3.74\times10^{-44}$      &       $1.50\times10^{-43}$      &       $3.79\times10^{-45}$      &       $1.52\times10^{-44}$ \\ \hline
2.7      &       $1.14\times10^{-43}$      &       $4.56\times10^{-43}$      &       $1.10\times10^{-44}$      &       $4.39\times10^{-44}$ \\ \hline
2.9      &       $2.66\times10^{-43}$      &       $1.07\times10^{-42}$      &       $2.48\times10^{-44}$      &       $9.92\times10^{-44}$ \\ \hline
3.1      &       $5.27\times10^{-43}$      &       $2.11\times10^{-42}$      &       $4.80\times10^{-44}$      &       $1.92\times10^{-43}$ \\ \hline
3.3      &       $9.37\times10^{-43}$      &       $3.75\times10^{-42}$      &       $8.39\times10^{-44}$      &       $3.36\times10^{-43}$ \\ \hline
3.5      &       $1.54\times10^{-42}$      &       $6.15\times10^{-42}$      &       $1.36\times10^{-43}$      &       $5.44\times10^{-43}$ \\ \hline
3.7      &       $2.38\times10^{-42}$      &       $9.51\times10^{-42}$      &       $2.08\times10^{-43}$      &       $8.33\times10^{-43}$ \\ \hline
3.9      &       $3.50\times10^{-42}$      &       $1.40\times10^{-41}$      &       $3.05\times10^{-43}$      &       $1.22\times10^{-42}$ \\ \hline
4.1      &       $4.97\times10^{-42}$      &       $1.99\times10^{-41}$      &       $4.30\times10^{-43}$      &       $1.72\times10^{-42}$ \\ \hline
4.3      &       $6.86\times10^{-42}$      &       $2.74\times10^{-41}$      &       $5.87\times10^{-43}$      &       $2.35\times10^{-42}$ \\ \hline
4.5      &       $9.22\times10^{-42}$      &       $3.69\times10^{-41}$      &       $7.84\times10^{-43}$      &       $3.14\times10^{-42}$ \\ \hline
4.7      &       $1.21\times10^{-41}$      &       $4.85\times10^{-41}$      &       $1.03\times10^{-42}$      &       $4.11\times10^{-42}$ \\ \hline
4.9      &       $1.57\times10^{-41}$      &       $6.27\times10^{-41}$      &       $1.32\times10^{-42}$      &       $5.28\times10^{-42}$ \\ \hline
5.1      &       $1.98\times10^{-41}$      &       $7.94\times10^{-41}$      &       $1.66\times10^{-42}$      &       $6.65\times10^{-42}$ \\
\hline
\end{tabular}
%\end{table}
%
%\begin{table}[htbp]
\centering
\caption{Total cross sections (in $\mathrm{GeV}^{-2}$) for
$n\Sigma^0\to pp e^-e^-$ and $\Lambda\Lambda\to pp e^-e^-$ mediated by
heavy Majorana neutrino exchange.  The  parameters
$\eta_N^A$ and $\eta_N^B$ are defined in Tab.~\ref{tab:HN_NN_NL}. The
$\eta_N^B$ results are four times the corresponding $\eta_N^A$ results.}
\label{tab:HN_NS_LL}
\renewcommand{\arraystretch}{1.1}
\setlength{\tabcolsep}{25pt}
\begin{tabular}{c|cc|cc}
\hline\hline
\rule[-1.2ex]{0pt}{5.0ex}
\raisebox{-2.4ex}{\shortstack{$\sqrt{s}$\\{(GeV)}}}
     &       \multicolumn{2}{c|}{$n\Sigma^0\to pp e^-e^-$}
     &       \multicolumn{2}{c}{$\Lambda\Lambda\to pp e^-e^-$} \\
\cline{2-5}
&       $\eta_N^A$      &       $\eta_N^B$      &       $\eta_N^A$      &       $\eta_N^B$ \\\hline
2.15     &       $7.13\times10^{-48}$      &       $2.85\times10^{-47}$      &       --      &       -- \\ \hline
2.3      &       $4.67\times10^{-47}$      &       $1.87\times10^{-46}$      &       $1.26\times10^{-46}$      &       $5.04\times10^{-46}$ \\ \hline
2.5      &       $2.92\times10^{-46}$      &       $1.17\times10^{-45}$      &       $3.96\times10^{-46}$      &       $1.59\times10^{-45}$ \\ \hline
2.7      &       $1.02\times10^{-45}$      &       $4.07\times10^{-45}$      &       $1.05\times10^{-45}$      &       $4.20\times10^{-45}$ \\ \hline
2.9      &       $2.60\times10^{-45}$      &       $1.04\times10^{-44}$      &       $2.28\times10^{-45}$      &       $9.12\times10^{-45}$ \\ \hline
3.1      &       $5.48\times10^{-45}$      &       $2.19\times10^{-44}$      &       $4.32\times10^{-45}$      &       $1.73\times10^{-44}$ \\ \hline
3.3      &       $1.02\times10^{-44}$      &       $4.07\times10^{-44}$      &       $7.40\times10^{-45}$      &       $2.96\times10^{-44}$ \\ \hline
3.5      &       $1.73\times10^{-44}$      &       $6.91\times10^{-44}$      &       $1.18\times10^{-44}$      &       $4.74\times10^{-44}$ \\ \hline
3.7      &       $2.73\times10^{-44}$      &       $1.09\times10^{-43}$      &       $1.80\times10^{-44}$      &       $7.18\times10^{-44}$ \\ \hline
3.9      &       $4.10\times10^{-44}$      &       $1.64\times10^{-43}$      &       $2.61\times10^{-44}$      &       $1.04\times10^{-43}$ \\ \hline
4.1      &       $5.90\times10^{-44}$      &       $2.36\times10^{-43}$      &       $3.64\times10^{-44}$      &       $1.46\times10^{-43}$ \\ \hline
4.3      &       $8.22\times10^{-44}$      &       $3.29\times10^{-43}$      &       $4.97\times10^{-44}$      &       $1.99\times10^{-43}$ \\ \hline
4.5      &       $1.11\times10^{-43}$      &       $4.46\times10^{-43}$      &       $6.60\times10^{-44}$      &       $2.64\times10^{-43}$ \\ \hline
4.7      &       $1.47\times10^{-43}$      &       $5.89\times10^{-43}$      &       $8.60\times10^{-44}$      &       $3.44\times10^{-43}$ \\ \hline
4.9      &       $1.91\times10^{-43}$      &       $7.63\times10^{-43}$      &       $1.10\times10^{-43}$      &       $4.40\times10^{-43}$ \\ \hline
5.1      &       $2.42\times10^{-43}$      &       $9.69\times10^{-43}$      &       $1.39\times10^{-43}$      &       $5.55\times10^{-43}$ \\
\hline
\end{tabular}
\end{table}
\begin{table}[htbp]
\centering
\caption{Total cross sections (in $\mathrm{GeV}^{-2}$) for $\Sigma^0\Sigma^0\to pp e^-e^-$ and $\Lambda\Sigma^0\to pp e^-e^-$ mediated by heavy Majorana neutrino exchange.
The  parameters $\eta_N^A$ and $\eta_N^B$ are defined in Tab.~\ref{tab:HN_NN_NL}. The $\eta_N^B$ results are four times the corresponding $\eta_N^A$ results. }
\label{tab:HN_SL_SS}
\renewcommand{\arraystretch}{1.25}
\setlength{\tabcolsep}{25pt}
\begin{tabular}{c|cc|cc}
\hline\hline
\rule[-1.2ex]{0pt}{5.0ex}
\raisebox{-2.4ex}{\shortstack{$\sqrt{s}$\\{(GeV)}}}
     &       \multicolumn{2}{c|}{$\Sigma^0\Sigma^0\to pp e^-e^-$}
     &       \multicolumn{2}{c}{$\Lambda\Sigma^0\to pp e^-e^-$} \\
\cline{2-5}     &       $\eta_N^A$      &       $\eta_N^B$      &       $\eta_N^A$      &       $\eta_N^B$ \\\hline
2.32  &       --                        &       --                      &       $6.79\times10^{-48}$      &       $2.72\times10^{-47}$     \\ \hline
2.39  &       $5.85\times10^{-47}$      &       $2.34\times10^{-46}$    &         $\cdots$                &       $\cdots$                 \\ \hline
2.5   &       $2.98\times10^{-47}$      &       $1.19\times10^{-46}$    &       $2.02\times10^{-47}$      &       $8.08\times10^{-47}$      \\ \hline
2.7   &       $6.45\times10^{-47}$      &       $2.58\times10^{-46}$    &       $7.20\times10^{-47}$      &       $2.88\times10^{-46}$      \\ \hline
2.9   &       $1.33\times10^{-46}$      &       $5.31\times10^{-46}$    &       $1.87\times10^{-46}$      &       $7.47\times10^{-46}$      \\ \hline
3.1   &       $2.46\times10^{-46}$      &       $9.85\times10^{-46}$    &       $3.98\times10^{-46}$      &       $1.59\times10^{-45}$      \\ \hline
3.3   &       $4.19\times10^{-46}$      &       $1.67\times10^{-45}$    &       $7.44\times10^{-46}$      &       $2.98\times10^{-45}$      \\ \hline
3.5   &       $6.66\times10^{-46}$      &       $2.66\times10^{-45}$    &       $1.27\times10^{-45}$      &       $5.07\times10^{-45}$      \\ \hline
3.7   &       $1.00\times10^{-45}$      &       $4.02\times10^{-45}$    &       $2.01\times10^{-45}$      &       $8.03\times10^{-45}$      \\ \hline
3.9   &       $1.45\times10^{-45}$      &       $5.82\times10^{-45}$    &       $3.01\times10^{-45}$      &       $1.21\times10^{-44}$      \\ \hline
4.1   &       $2.02\times10^{-45}$      &       $8.08\times10^{-45}$    &       $4.36\times10^{-45}$      &       $1.74\times10^{-44}$      \\ \hline
4.3   &       $2.74\times10^{-45}$      &       $1.10\times10^{-44}$    &       $6.07\times10^{-45}$      &       $2.43\times10^{-44}$      \\ \hline
4.5   &       $3.64\times10^{-45}$      &       $1.46\times10^{-44}$    &       $8.24\times10^{-45}$      &       $3.29\times10^{-44}$      \\ \hline
4.7   &       $4.73\times10^{-45}$      &       $1.89\times10^{-44}$    &       $1.09\times10^{-44}$      &       $4.35\times10^{-44}$      \\ \hline
4.9   &       $6.03\times10^{-45}$      &       $2.41\times10^{-44}$    &       $1.41\times10^{-44}$      &       $5.63\times10^{-44}$      \\ \hline
5.1   &       $7.55\times10^{-45}$      &       $3.02\times10^{-44}$    &       $1.79\times10^{-44}$      &       $7.16\times10^{-44}$      \\ \hline
\end{tabular}
\end{table}

 \end{appendix}

\newpage

\end{document}